\documentclass[a4paper, 10pt]{report}

\usepackage[cp1250]{inputenc}
\usepackage[T1]{fontenc}
\usepackage{amsfonts}%toto je na symbol reálnych čísel
\usepackage{graphicx}
\usepackage{amsmath}
\usepackage{caption}

\usepackage{amssymb}%advances AMS symbols

\usepackage{fancyhdr}
\usepackage{bibtopic}
\usepackage{color}
\usepackage{ulem}
\usepackage[toc,page]{appendix}
\usepackage{setspace}
\usepackage{cite}
\usepackage{xcolor}

\AtBeginDocument{}
\usepackage{etoolbox}
\patchcmd{\thebibliography}{\chapter*}{\section*}{}{}

\makeatletter
\renewcommand{\thesection}{%
  \ifnum\c@chapter<1 \@arabic\c@section
  \else \thechapter.\@arabic\c@section
  \fi
}
\makeatother

\patchcmd{\tableofcontents}{\chapter*}{\section*}{}{}

\numberwithin{equation}{section}

\let\OLDthebibliography\thebibliography
\renewcommand\thebibliography[1]{
  \OLDthebibliography{#1}
  \setlength{\parskip}{0pt}
  \setlength{\itemsep}{3.5pt plus 1ex}
}

\usepackage[linktocpage]{hyperref}
\definecolor{darkred}{rgb}{0.5,0,0}
\definecolor{darkpurple}{rgb}{0.5,0,0.5}
\definecolor{darkblue}{rgb}{0,0,0.5}
\hypersetup{ colorlinks,
linkcolor=darkblue,
filecolor=darkpurple,
urlcolor=darkred,
citecolor=darkpurple }
 
\begin{document}

\allowdisplaybreaks
\setlength{\abovedisplayskip}{3.5pt}
\setlength{\belowdisplayskip}{3.5pt}
\abovedisplayshortskip
\belowdisplayshortskip

{\setstretch{1.0}

{\LARGE \bf \centerline{
Inflationary tensor fossils deformed by
}}
{\LARGE \bf \centerline{
solid matter - scalar field interaction
}}

\vskip 1cm
\begin{center}
{Peter M\'esz\'aros}

\vskip 2mm {\it Department of Theoretical Physics, Comenius
University, Bratislava, Slovakia}

\vskip 2mm \today %May 31, 2019
\end{center}

\section*{Abstract}

\noindent We study effect of superhorizon tensor perturbations on scalar perturbations, so called effect of clustering fossils, in cosmological model in which inflation is driven by both solid matter and scalar field. The effect deforms primordial scalar power spectrum causing asymmetry in it, which leaves imprint on CMB anisotropies and the cosmic structure. Parameter space of this combined model allows for enhancement of logarithmic scale dependence of this deformation, as opposed to simpler models in which such scale dependence is suppressed by slow-roll parameters.

\vskip 3mm \hspace{4mm}
\begin{minipage}[t]{0.8\textwidth}
\noindent\rule{12cm}{0.4pt}
\vspace{-9mm}
\tableofcontents
\noindent\rule{12cm}{0.4pt}
\end{minipage}
\vskip 6mm

}

\section{Introduction}\label{sec:1}

Models predicting inflationary expansion in very early universe and their variations are challenged by increase in accuracy of observations of cosmic microwave background (CMB) anisotropies \cite{planck18}. One of the most important areas of theoretical predictions which still lacks direct observational evidence concerns tensor perturbations or primordial gravitational waves. Data obtained by their observations may either rule out some models or considerably restrict their parameter spaces. Currently available observational restriction on tensor perturbations comes from the fact that they still have not been detected, giving upper limit on tensor-to-scalar ratio, $r < 0.064$ \cite{planck18, bicep}.

\vskip 1mm
In recent years an interesting consequence of presence of tensor perturbations in the early universe, so called effect of clustering fossils, was studied in context of various inflationary models. The effect is based on nonlinearity of the perturbation theory due to which evolution of perturbations is influenced by long wavelength modes frozen on superhorizon scale \cite{jeong, dai}. In particular, superhorizon tensor perturbations affect subhorizon scalar perturbations in such way that their scalar power spectrum obtains quadrupole asymmetry. This leaves imprint on CMB anisotropies, and consequently also on nonlinear structures formed in later stages of evolution of the universe, which can be observed in galaxy surveys \cite{brahma, dimastrogiovanni, emami}.

\vskip 1mm
In the single-field model, if we disregard non-Einstein theories of gravity, the nonlinearity of perturbations arises only from nonlinear structure of Einstein--Hilbert action, and therefore, it is weak \cite{maldacena}. Three-point correlation functions in this model then satisfy consistency relations \cite{creminelli, berezhiani}, and the quadrupole asymmetry is cancelled out with effect of tensor perturbations on light travelling to the observer \cite{book, dai2}. On the other hand, from the point of view of nonlinear effects multi-field models are much more interesting. Among such models an inflationary model in which consistency relations are broken and with unique prediction for the scalar bispectrum is so called solid inflation \cite{gruzinov, endlich}. It is described by three fields which enter the matter Lagrangian in the same way as body coordinates of a homogeneous and isotropic elastic matter enter its equation of state. The predicted bispectrum peaks in the squeezed limit as in other multi-field models, but its unique feature is an anisotropic dependence of sign of the peak on direction in which the limit is approached. Within this model the effect of clustering fossils was studied in \cite{dimastrogiovanni, akhshik}, and the model was also further generalized by authors of \cite{bartolo, rew, ricciardone, ja}.

\vskip 1mm
In this paper we consider a simple generalization of solid inflation by adding a single scalar field with standard kinetic term to it. In this way we can study interaction of two kinds of matter of different nature during inflationary expansion driven by them. The model can be also considered as a simple toy model for interaction between fields driving the solid inflation and fields of the standard model. The main result of the work is that the quadrupole deformation of the primordial scalar power spectrum is allowed to have an unusually large scale dependence, whereas in simpler models this dependence is mild, suppressed by slow-roll parameters, and therefore, it is usually disregarded. The model studied in this paper shares this feature with an inflationary model studied in \cite{rew}, where the scalar field added to three fields breaking spatial diffemorphism symmetry breaks time diffeomorphism symmetry generalizing the diffeomorphism breaking pattern.

\vskip 1mm
The following text is divided into two main parts. In the first part, in section \ref{sec:2}, we summarize main results of the previous work studying a combined inflationary model with both solid matter and scalar field \cite{ja}. For simplicity, we mostly restrict ourselves only to the lowest order of the slow-roll approximation, but we present restrictions on parameters of the model given by slow-roll approximation in a more systematic way, see also appendix \ref{app:a}, and in addition to the old results we derive relation between the local nonlinearity parameter and angle of direction in which the squeezed limit is approached. Considering the next to the leading order of the slow-roll approximation we also calculate the scalar spectral tilt. In the second part of the paper, in section \ref{sec:3}, we provide the analysis of clustering fossils within the model, and we discuss the result. There is also a concluding section \ref{sec:4}, and two more appendices including additional materials. We use units in which $\hbar = c = 1$ and signature $(+,-,-,-)$ for the space-time metric.

\section{Model with scalar field and solid matter}\label{sec:2}

There are two different kinds of fields entering the matter Lagrangian of the model. The first one is the scalar field which is a function from space-time to real axis, $\varphi: (\mathcal{M}^{(1,3)}, g) \to (\mathbb{R}, 2 \mathcal{K})$, with kinetic term $\mathcal K$ defined as a push-forward of the space-time metric with respect to this function multiplied by factor $1/2$, i.e. $\mathcal{K} = \varphi_{*} g / 2$, or in coordinates
\begin{eqnarray}
\mathcal{K} %= \frac{1}{2} g_{\mu \nu} \partial^\mu \varphi \partial^\nu \varphi
= \frac{1}{2} g^{\mu \nu} \varphi_{,\mu} \varphi_{,\nu}.
\end{eqnarray}
The second kind of fields corresponds to function from space-time to 3-dimensional manifold, $\Phi: (\mathcal{M}^{(1,3)}, g) \to (\mathcal{N}^{(3)}, -B)$, which can be interpreted as a body manifold associated with continuum or solid matter. Denote coordinates on the manifold $\mathcal{N}^{(3)}$, so called body coordinates, as $\phi^I$, $I = 1, 2, 3$, and describe the mapping with use of them. Push-forward of the space-time metric $\Phi_{*} g \equiv -B$ then defines components of body metric $B$ as
\begin{eqnarray}
\label{eq:bodymetric}
B^{IJ} = - g^{\mu\nu} \phi^I_{\phantom{I},\mu} \phi^J_{\phantom{I},\nu},
\end{eqnarray}
where the sign is changed in order to obtain positive metric $B$ for the signature which we are using.

\vskip 1mm
The body metric $B$ completely describes deformation of the solid, so that it is reasonable to consider its energy density, or part of the matter Lagrangian $\mathcal{L}_{\textrm{sol}}$ corresponding to it, to be a function of $B_{IJ}$. Furthermore, if we restrict ourselves to the case of homogeneous and isotropic solid, we have to demand invariance of the action with respect to global translations and rotations in the body space,
\begin{eqnarray}
\label{eq:symmetry}
\phi^I \mapsto R^{I}_{\phantom{I}J} \phi^J + T^I, \qquad R^{I}_{\phantom{I}J} \in SO(3), \qquad T^{I} \in \mathbb{R}^3.
\end{eqnarray}
In other words, the energy density of the solid depends only on its pure deformations form which global translations and rotations are excluded. This is a key property of solids which is taken into account when Newtonian theory of elastic continuum is built. The solid matter Lagrangian density with this symmetry can be parametrized by invariant quantities defined through the body metric. There are at most three such invariants which are independent on each other, and we are choosing them in the same way as in \cite{endlich},
\begin{eqnarray}
\label{eq:invariants}
X = \textrm{Tr} B, \qquad Y = \frac{\textrm{Tr} B^2}{X^2}, \qquad Z = \frac{\textrm{Tr} B^3}{X^3}.
\end{eqnarray}
The solid matter Lagrangian density written as $\mathcal{L}_{\textrm{sol}}=\mathcal{E}(X,Y,Z)$ is then invariant with respect to transformations (\ref{eq:symmetry}) automatically. Relation between this description of relativistic elasticity and its Newtonian limit described by Lam\'e parameters can be found in appendix \ref{app:b}.

\vskip 1mm
The simplest matter Lagrangian depending on both scalar field and homogeneous and isotropic solid matter then can be written as $\mathcal{L}_{\textrm{mat}} = \mathcal{K} - V(\varphi) + \mathcal{E}(X,Y,Z)$, where $V$ is potential of the scalar field. With this choice any non-trivial coupling between the two matter components is disregarded and coupling arises only as a effect of gravity added to the theory by including the Einstein--Hilbert Lagrangian density into the overall action. However, in order to have a model with non-trivial coupling between scalar field and solid we consider a more general form of the matter Lagrangian density, $\mathcal{L}_{\textrm{mat}} = \mathcal{K} + F(\varphi,X,Y,Z)$. The overall action which defines the model is then
\begin{eqnarray}
\label{eq:action}
S = \int{\sqrt{-g} d^4x \left(\frac{1}{2} M_{\textrm{Pl}}^2 R + \mathcal{K} + F(\varphi,X,Y,Z)\right)},
\end{eqnarray}
with Ricci scalar denoted by $R$ and $M_{\textrm{Pl}}$ being the reduced Planck mass. With this choice we avoid non-trivial coupling of gravity to the matter, which was studied by authors of \cite{ricciardone} in the context of solid inflation.

\vskip 1mm
Note also that coupling between scalar field and solid matter could be generalized further by including dependence on $w^I = g^{\mu\nu} \varphi_{,\mu} \phi^I_{\phantom{I},\mu}$. To preserve symmetry with respect to internal transformations (\ref{eq:symmetry}) the matter Lagrangian would explicitly depend on invariant quantities constructed from $w^I$, for example determinant or trace of matrix $w^I w^J$ and its powers. However, with this generalization it is no longer natural to describe the theory by functions $\varphi$ and $\Phi$, but rather by a function $\Psi: (\mathcal{M}^{(1,3)}, g) \to (\mathcal{N}^{(4)}, \Psi_{*}g)$, where $SO(3)$ group describes symmetry of the 4-dimensional manifold $\mathcal{N}^{(4)}$, so that this manifold is anisotropic. Coordinates $(\varphi, \phi^I)$ are then chosen in such way that coordinates $\phi^I$ measure positions on isotropic hypersurfaces in $\mathcal{N}^{(4)}$ and coordinate $\varphi$ labels these hypersurfaces. Authors of \cite{rew} consider a 4-component field corresponding to function $\Psi$ and set its background configuration to coincide with space-time coordinates, cosmic time and comoving spatial coordinates.

\subsection{Slow-roll inflation}

There is a variety of cosmological solutions associated with action (\ref{eq:action}), depending on function $F(\varphi,X,Y,Z)$ and initial conditions. Of course, majority of them cannot be expected to be applicable to our universe, but some of them can be, for instance solutions corresponding to cosmological inflation.

\vskip 1mm
When studying the inflation within this model we restrict ourselves to case of flat homogeneous and isotropic universe. The space-time metric is then the metric of flat Friedmann--Robertson--Walker--Lema\^{i}tre (FRWL) universe,
\begin{eqnarray}
\label{eq:FRWL}
ds^2 = dt^2 - a^2(t) \delta_{ij} dx^i dx^j,
\end{eqnarray}
where $a$ is the scale factor, $t \equiv x^0$ is the cosmic time and $x^i$ denote comoving spatial coordinates. The requirement of homogeneity and isotropy is satisfied when the scalar field $\varphi$ does not depend on space coordinates, $\varphi = \varphi(t)$, and the body coordinates fields coincide with comoving spatial coordinates,
\begin{eqnarray}
\label{eq:phi0}
\phi^I = \delta^I_i x^i.
\end{eqnarray}
This configuration breaks rotational and translational symmetry, but it is invariant under combined spatial-internal translations and rotations. Choosing fields $\phi^I$ to be functions of only the time, similarly as for the scalar field, would preserve homogeneity and isotropy as well, but with this choice the solid matter would define a privileged frame coupled to it which differs from frame coupled to the rest of matter filling the universe. By using (\ref{eq:bodymetric}), (\ref{eq:FRWL}) and (\ref{eq:phi0})  invariants (\ref{eq:invariants}) in this homogeneous and isotropic case can be expressed as
\begin{eqnarray}
\label{eq:Xbar}
X = \frac{3}{a^2}, \qquad Y = \frac{1}{3}, \qquad Z = \frac{1}{9}.
\end{eqnarray}
Notice that only the first of them explicitly depends on time.

\vskip 1mm
The action (\ref{eq:action}) determines equations for homogeneous and isotropic theory, which is parametrized by two independent functions of time, $a(t)$ and $\varphi(t)$. The system of equations is
\begin{eqnarray}
\label{eq:friedmann1}
\dot{\varphi}^2 - 6 M_{\textrm{Pl}}^2 H^2 & = & 2 F, \\
\label{eq:friedmann2}
\dot{\varphi}^2 + 2 M_{\textrm{Pl}}^2 \dot{H} & = & 2 a^{-2} F_X, \\
\label{eq:kleingordon}
\ddot{\varphi} + 3 H \dot{\varphi} & = & F_{\varphi},
\end{eqnarray}
with $H = \dot{a} / a$ denoting the Hubble parameter and subscripts standing for partial derivatives of the function $F$. The first two equations are Friedmann equations, the third one generalizes Klein--Gordon equation for the scalar field to the case of flat FRWL universe, and due to the Bianchi identity it is not independent from the first two ones. In order to obtain slow-roll inflation as a solution of this system of equations, Hubble flow parameters
\begin{eqnarray}
\label{eq:slowroll}
\epsilon \equiv \eta^{(0)} = - \frac{\dot{H}}{H^2}, \qquad \eta \equiv \eta^{(1)}, \qquad \eta^{(n+1)} = \frac{d \ln \eta^{(n)}}{d \ln a} = \frac{\dot{\eta}^{(n)}}{H \eta^{(n)}},
\end{eqnarray}
must be small. In such case they are called slow-roll parameters. With the use of equations (\ref{eq:friedmann1}) and (\ref{eq:friedmann2}) the slow-roll parameter $\epsilon$ can be expressed as
\begin{eqnarray}
\label{eq:epsilon}
\epsilon = p + q - \frac{1}{3} p q, \qquad p = \frac{\dot{\varphi}^2}{2 M_{\textrm{Pl}}^2 H^2}, \qquad q = X \frac{F_X}{F},
\end{eqnarray} 
where $p$ and $q$ are of the same form as slow-roll parameters of model in which the inflation is driven only by scalar field and only by solid matter respectively. Of course, in our combined model $p$ and $q$, in principle, may be not small, as long as their combination $\epsilon$, which measures the Hubble flow, is small. However, stability of scalar perturbations, which will be discussed in what follows, requires smallness of both of these parameters.

\vskip 1mm
As we can see, not every function $F(\varphi,X,Y,Z)$ is compatible with slow-roll restrictions. From (\ref{eq:epsilon}) it is obvious that there are restrictions at least on first partial derivatives of $F$. Actually, this concerns also higher derivatives, and restrictions on them will be specified in the following subsection.

\subsection{Cosmological perturbations}

We summarize results of the perturbation theory for inflationary model defined above. In order to avoid complication arising from interaction between two matter components, we restrict ourselves to only the leading order of the slow-roll approximation applied to the case with supressed coupling between scalar field and solid matter, see (\ref{eq:weakcoupling}). In this way, we present size of the scalar and tensor power spectra and tensor-to-scalar ratio, and in subsection \ref{0000} we consider also the next to leading order of the slow-roll approximation and calculate the scalar spectral tilt. The scalar bispectrum will be included in the next subsection.

\vskip 1mm
Perturbation theory in the model where the role of the background solution is played by the homogeneous and isotropic case discussed above is built by adding perturbations to space-time metric as well as matter components. It is convenient to employ Arnowitt--Deser--Misner (ADM) formalism, with space-time metric parametrised by
\begin{eqnarray}
ds^2 = N^2 dt^2 - h_{ij} (dx^i + N^i dt) (dx^j + N^j dt),
\end{eqnarray}
because we can make use of momentum and Hamiltonian constraints. The flat FRWL metric corresponds to $N = 1$, $N^i = 0$ and $h_{ij} = a^2 \delta_{ij}$, and we define its perturbations by quantities $\delta N$, $\xi$, $N^i_{\perp}$ and $\gamma_{ij}$ as
\begin{eqnarray}
\label{eq:metricperturbations}
N = 1 + \delta N, \qquad N^i = \xi_{,i} + N^i_{\perp}, \qquad h_{ij} = a^2 \exp (\gamma_{ij}),
\end{eqnarray}
where $N^i_{\perp,i} = 0$, so that $N^i$ is decomposed into its scalar and vector parts and tensor perturbation $\gamma_{ij}$ is traceless, $\gamma_{ii} = 0$, and transversal, $\gamma_{ij,j} = 0$. With such parametrisation of metric perturbations we have set spatially flat slicing gauge with scalar and vector perturbations of the 3-dimensional metric $h_{ij}$ set to zero. With all the gauge freedom used up for perturbations of the metric, perturbations of all matter fields must be taken into account,
\begin{eqnarray}
\label{eq:displacement}
\varphi = \overline{\varphi} + \delta\varphi, \qquad \phi^I = \overline{\phi}^I + \delta^{Ii} \rho_{,i} + \pi^I_{\perp},
\end{eqnarray}
with bar denoting fields of the unperturbed theory and vector part of the perturbation $\pi^I$ satisfying the decomposition condition $\delta^i_I \pi^I_{\perp,i} = 0$. Vector and tensor perturbations can be further decomposed into independent polarizations,
\begin{eqnarray}
\delta^i_I \pi^I_{\perp \mathbf{k}} = \sum\limits_{P = \pm} \varepsilon^i_{P \mathbf{k}} \pi_{\perp P \mathbf{k}}, \qquad \gamma_{\mathbf{k} i j} = \sum\limits_{\lambda = + , \times} e^{\lambda}_{\mathbf{k} i j} \gamma^{\lambda}_{\mathbf{k}},
\end{eqnarray}
where the polarization vectors satisfy the transverse condition, $k_i \varepsilon^i_{\mathbf{k}} = 0$, as well as the polarization tensors, $k_i e_{\mathbf{k} ij} = 0$, which satisfy also the traceless condition, $e_{\mathbf{k} i i} = 0$. As a normalization conditions for them we use relations $\varepsilon^i_{P \mathbf{k}} \varepsilon^{i*}_{P^\prime \mathbf{k}} = 2 \delta_{P P^\prime}$ and $e^{\lambda}_{\mathbf{k} ij} e^{\lambda^\prime *}_{\mathbf{k} ij} = 2 \delta_{\lambda \lambda^\prime}$.

\vskip 1mm
Convenience of decomposition of perturbations into scalar, vector and tensor perturbations manifests in decomposability of the quadratic action into three parts, the scalar quadratic action $S^{(2)}_{\textrm{S}}$, vector quadratic action $S^{(2)}_{\textrm{V}}$, and tensor quadratic action $S^{(2)}_{\textrm{T}}$, $S^{(2)} = S^{(2)}_{\textrm{S}} + S^{(2)}_{\textrm{V}} + S^{(2)}_{\textrm{T}}$. In order to obtain these actions, the action (\ref{eq:action}) must be expanded up to the second order. From the first order part of the expansion we obtain momentum and Hamiltonian constraints which must be satisfied. As a result, there remain only four independent perturbations, two scalar perturbations $\delta\varphi$ and $\rho$, vector perturbation $\pi^I_{\perp}$ and tensor perturbation $\gamma_{ij}$. Quadratic actions expressed in terms of them are
\begin{eqnarray}
\label{eq:scalar}
S^{(2)}_{\textrm{S}} & = & \int \frac{d t d^3 k}{(2 \pi)^3} a^3 \bigg[ M_{\textrm{Pl}}^2 \mathcal{Q} k^4 \left| \dot{\rho} \right|^2 + \frac{1}{2} \left( 1 + p\mathcal{Q} \right) \left| \dot{\delta\varphi} \right|^2 + \\
& & + \phantom{00} \textrm{non-kinetic terms including} \phantom{00} - M_{\textrm{Pl}}^2 H^2 Q \left( c_L^2 - Q \mathcal{Q} \right) k^4 \left| \rho \right|^2 \bigg], \nonumber \\
\label{eq:vector}
S^{(2)}_{\textrm{V}} & = & M_{\textrm{Pl}}^2 \int \frac{d t d^3 k}{(2 \pi)^3} a^3 \left( \frac{a^2 H^2 Q k^2}{4 a^2 H^2 Q + k^2} \left| \dot{\pi}^I_{\perp} \right|^2 - H^2 Q c_T^2 k^2 \left| \pi^I_{\perp} \right|^2 \right), \\
\label{eq:tensorqqq}
S^{(2)}_{\textrm{T}} & = & \frac{M_{\textrm{Pl}}^2}{4} \int d^4 x a^3 \left( \frac{1}{2} \left| \dot{\gamma}_{ij} \right|^2 - \frac{1}{2} a^{-2} \left| \gamma_{ij,k} \right|^2 - 2 H^2 Q c_T^2 \left| \gamma_{ij} \right|^2 \right),
\end{eqnarray}
where $Q = \epsilon - p$, $\mathcal{Q} = a^2 H^2 Q / \left[ a^2 H^2 \left( 3 - p \right) Q + k^2 \right]$, and $c_L$ and $c_T$ denote longitudinal and transverse sound speed respectively. Their squares can be expressed in terms of the function $F$ and its derivatives as
\begin{eqnarray}
\label{eq:soundspeeds}
c_L^2 = 1 + \frac{2}{3} \frac{\overline{X} \overline{F}_{XX}}{\overline{F}_X} + \frac{8}{9} \frac{\overline{F}_Y + \overline{F}_Z}{\overline{X} \overline{F}_X}, \qquad
c_T^2 = 1 + \frac{2}{3} \frac{\overline{F}_Y + \overline{F}_Z}{\overline{X} \overline{F}_X},
\end{eqnarray}
with overline denoting quantities evaluated at the backgrond configuration (\ref{eq:Xbar}), and for simplicity, from now on we will omit it. We have not written down most of terms in the scalar quadratic action (\ref{eq:scalar}) and we focused on kinetic terms and the term with longitudinal sound speed. Importance of kinetic terms lies in the fact that in order to avoid instability of scalar perturbations they must have the proper sign. As a consequence, both $Q$ and $\mathcal{Q}$ must be positive, and therefore, slow-roll parameters must satisfy $0 < p < \epsilon$, i.e. not only $\epsilon$ but also both $p$ and $q$ must be small. Moreover, we will consider smallness of parameters $\eta_Q$, $\eta^{(2)}_Q$, ... defined by (\ref{eq:slowroll}) with $\epsilon$ replaced by $Q$, $\eta_L$, $\eta^{(2)}_L$, ... defined in the same way by the longitudinal sound speed, $\eta_T$, $\eta^{(2)}_T$, ... by the transverse sound speed and $\eta_p$, $\eta^{(2)}_p$, ... $\eta_q$, $\eta^{(2)}_q$, ... defined by $p$ and $q$. This choice is motivated by simplifying the analysis of cosmological perturbations and avoiding too large scalar spectral tilt calculated in subsection \ref{0000}.

\vskip 1mm
As we can see, in order to avoid superluminal propagation of perturbations, derivatives of function $F$ appearing in relation (\ref{eq:soundspeeds}) must be restricted. In addition to this restriction we have to take into account also smallness of all slow-roll parameters defined so far. With the use of their definitions together with equations of motion (\ref{eq:friedmann1})-(\ref{eq:kleingordon}) we find that perturbative expansion of function $F$ up to the second order can be expressed in terms of sound speeds and slow-roll parameters,
\begin{eqnarray}
\label{eq:Fexpand}
F & = & M_{\textrm{Pl}}^2 H^2 (-3 + p) \bigg[ 1 + q \frac{\delta X}{X} \mp \sqrt{2p} \frac{6 - 2\epsilon + \eta_p}{6 - 2p} \frac{\delta \varphi}{M_{\textrm{Pl}}} - \\
& & - \mathcal{C}_1 q \frac{\delta X^2}{X^2} + \frac{F_Y}{F} \delta Y - \left( \frac{3}{2} \mathcal{C}_2 q + \frac{F_Y}{F} \right) \delta Z \mp \nonumber\\
& & \mp 3 \sqrt{2} \mathcal{C}_3 \frac{q}{\sqrt{p}} \frac{\delta X}{X} \frac{\delta \varphi}{M_{\textrm{Pl}}} - 3 \mathcal{C}_4 \frac{q}{p} \frac{\delta \varphi^2}{M_{\textrm{Pl}}^2} + \mathcal{O} \left( \chi^3 \right)
\bigg], \nonumber
\end{eqnarray}
with $\chi$ denoting any of considered perturbations or linear combinations of them. There are restrictions also on higher order terms, for more details see appendix \ref{app:a}. We can see that $F_Y/F$ is the only parameter of the theory which is not restricted by slow-roll approximation, if the perturbation theory is considered only up to the second order. Note that $\delta Y$ and $\delta Z$ are of the second order in the perturbation theory, absolute values of parameters $\mathcal{C}_1$ and $\mathcal{C}_2$ are not larger than one, and the same is true for $\mathcal{C}_3$ and $\mathcal{C}_4$ for slow-roll parameters set to zero. They can be expressed as
\begin{eqnarray}
\label{eq:cecka}
& & \mathcal{C}_1 = c_T^2 - \frac{3}{4} c_L^2 - \frac{1}{4}, \qquad \mathcal{C}_2 = 1 - c_T^2, \qquad  \mathcal{C}_3 = \frac{2}{3} \mathcal{C}_1 - \frac{1}{3} \left(1 - \epsilon + \frac{1}{2} \eta_Q \right), \\
& & \mathcal{C}_4 = \mathcal{C}_3 + \frac{p}{18 Q} \left[ \left( - 2 \epsilon + \frac{1}{2} \eta_p \right) \left( 3 - \epsilon + \frac{1}{2} \eta_p \right) + \frac{1}{2} \eta_p \eta^{(2)}_p - \epsilon \eta \right]. \nonumber
\end{eqnarray}

\vskip 1mm
With both $p$ and $q$ being small, the scalar quadratic action (\ref{eq:scalar}) in the leading order of slow-roll approximation reduces to
\begin{eqnarray}
\label{eq:reducedaction}
S^{(2)}_{\textrm{S}} & = & \int \frac{d t d^3 k}{(2 \pi)^3} a^3
\bigg[
M_{\textrm{Pl}}^2 H^2 Q a^2 k^2 \left( \left| \dot{\rho} \right|^2 - c_L^2 \frac{k^2}{a^2} \left| \rho \right|^2 \right) + \frac{1}{2} \left| \dot{\delta\varphi} \right|^2 + \\
& + & \frac{1}{2} \left( 18 H^2 \mathcal{C}_4 \frac{Q}{p} - \frac{k^2}{a^2} \right) \left| \delta \varphi \right|^2 \mp 6 \sqrt{2} M_{\textrm{Pl}} H^2 \mathcal{C}_3 \frac{Q}{\sqrt{p}} k^2 \textrm{Re}\left\{ \rho \delta \varphi^{*} \right\}
\bigg]. \nonumber
\end{eqnarray}
The simplest way how to avoid non-flatness of scalar power spectrum is considering the term with $\mathcal{C}_4$ as well as the coupling term with $\mathcal{C}_3$ to be negligible. They can be at most of the same order as slow-roll parameters and this requirement can be satisfied by two choices. The first one is that $Q = \epsilon - p$ is much smaller than other slow-roll parameters and the second choice is that $\mathcal{C}_1 = 1/2$ in the leading order of the slow-roll expansion. In what follows we restrict ourselves to the latter choice, because smallness of $Q$ leads to even larger enhancement of the primordial non-Gaussianity in the limit of small longitudinal sound speed $c_L$, which is an interesting case from the point of view of clustering fossils studied in section \ref{sec:3}. As a consequence of this choice the two sound speeds are not independent, and the constraint
\begin{eqnarray}
\label{eq:soundconstraint}
\frac{4}{3} c_T^2 - c_L^2 = 1 + \mathcal{O}\left( \epsilon \right) 
\end{eqnarray}
must be satisfied. The transverse sound speed $c_T$ then must be larger than factor $\sqrt{3}/2 + \mathcal{O} \left( \epsilon \right)$, and the longitudinal sound speed must be smaller than $1/\sqrt{3} + \mathcal{O} \left( \epsilon \right)$. Relation between the condition $Q \ll \epsilon$ and validity of constraint on sound speeds (\ref{eq:soundconstraint}) can be easily seen when the longitudinal sound speed is rewritten as
\begin{eqnarray}
c_L^2 = \frac{1}{3} - \frac{p}{9Q} \frac{F_{\varphi\varphi}}{H^2} + \frac{8}{9} \frac{F_Y + F_Z}{X F_Z} + \mathcal{O} \left( \epsilon \right).
\end{eqnarray}
The same constraint on sound speeds as (\ref{eq:soundconstraint}) appears in model in which inflation is driven only by solid matter, and the alternative way to avoid non-flat spectrum, choosing $Q \ll \epsilon$, means that the Hubble slow-roll parameter $\epsilon$ approximately equals the scalar field slow-roll parameter $p$, i.e. the dominant contribution to the total energy density dictating the inflationary expansion is given by the scalar field.

\vskip 1mm
Constraint (\ref{eq:soundconstraint}) is valid also in the case with no non-trivial coupling between solid matter and scalar field when the function $F$ reduces to form discussed above, $F(\varphi,X,Y,Z) = V(\varphi) + \mathcal{E}(X,Y,Z)$, so that $F_{\varphi X} = 0$. As we can see from perturbative expansion of the function $F$ (\ref{eq:Fexpand}), its partial derivatives are not independent, and $F_{\varphi\varphi} = \pm \sqrt{2/p} M_{\textrm{Pl}}^{-1} X F_{\varphi X} - 6H^2 \epsilon + \mathcal{O}(\epsilon^2)$. With no non-trivial coupling between two matter components the scalar field mass term is then suppressed by the slow-roll parameter which results in mildness of the scalar power spectrum tilt. For the model discussed in this paper it is not necessary to restrict ourselves to this special case with $F_{\varphi X}$ set to zero and it is sufficient to keep it small enough,
\begin{eqnarray}
\label{eq:weakcoupling}
\left| k^2 X F_{\varphi X} \right| \ll Q^{1/2}.
\end{eqnarray}
Equations of motion for perturbations can be obtained by varying the quadratic actions. In regime of weak coupling (\ref{eq:weakcoupling}), which allows for nearly flat scalar power spectrum, equations for all considered perturbations in the leading order of slow-roll approximation can be written in the form
\begin{eqnarray}
\label{eq:equationforperturbations}
\chi_k^{\prime\prime} + \frac{\alpha}{\tau} \chi_k^{\prime} + \beta^2 k^2 \chi_k = 0,
\end{eqnarray}
with prime denoting differentiation with respect to the conformal time $\tau$ defined in the standard way as $\tau = \int a^{-1} dt$. For slow-roll parameters set to zero solution of the background equations (\ref{eq:friedmann1})-(\ref{eq:kleingordon}) is the de Sitter universe, and for this solution it is conventional to choose integration constant in the definition of the conformal time in such way that $\tau \in (- \infty, 0)$. Equation for solid matter scalar perturbation is equation (\ref{eq:equationforperturbations}) with $\chi_k = \rho_k$, $\alpha = -4$ and $\beta = c_L$, for scalar field perturbation it is with $\chi_k = \delta \varphi_k$, $\alpha = -2$ and $\beta = 1$, for vector perturbations we have to use substitution $\chi_k = v_{\perp P k} = 4 H Q k^{-2} (-\tau)^{-1} \pi_{\perp P k}^{\prime}$, set $\alpha = -2$ and $\beta = c_T$, and for tensor perturbations $\chi_k = \gamma_k^{\lambda}$, $\alpha = -2$ and $\beta = 1$. Note that $\sum_P \varepsilon^i_{P k} v_{\perp P k}$ equals $N^i_{\perp k}$ defined in (\ref{eq:metricperturbations}) in the leading order of slow-roll approximation.

\vskip 1mm
Each perturbation with any polarisation, if it is relevant for it, can be quantised by decomposing it with respect to creation and annihilation operators, $\chi_{\mathbf{k}} = \chi^{(\textrm{cl})}_{\mathbf{k}} a_{\mathbf{k}} + \chi^{(\textrm{cl}) *}_{- \mathbf{k}} a^{\dagger}_{- \mathbf{k}}$, obeying standard commutation relations with the only nonzero being $\left[a_{\mathbf{k}}, a^{\dagger}_{\mathbf{k}^\prime}\right] = (2\pi)^3 \delta^{(3)} ( \mathbf{k} - \mathbf{k}^\prime )$, where both operators in the commutator correspond to the same perturbation with the same polarization. Classical modes $\chi^{(\textrm{cl})}$ are solutions of equation (\ref{eq:equationforperturbations}) normalised in such way that equal time commutation relations for perturbations $\chi$ and their conjugate momenta, $\left[\chi(\mathbf{x}_1,t),\Pi_{\chi}(\mathbf{x}_2,t)\right]=i\delta^{(3)}\left(\mathbf{x}_1-\mathbf{x}_2\right)$, with others being zero, are in agreement with canonical commutation relations for creation and annihilation operators. This normalisation can be satisfied by matching the canonically normalised fields, through which the kinetic Lagrangian density can be written as $\mathcal{L}_{\textrm{kin}}=(1/2)a\left|\dot{\chi}^{(\textrm{can})}\right|^2$, to free wave function of the Minkowski space $\left( 2 \beta k \right)^{-1/2}e^{-i \beta k \tau}$ with cosmic time replaced by the conformal time in the limit of very early time, $\tau \to - \infty$, when the modes are deep inside the horizon and space-time curvature does not effect their evolution.

\vskip 1mm
After solving equations for perturbations and employing the normalisation procedure explained above, we find the correctly normalised modes in the leading order of the slow-roll approximation,
\begin{eqnarray}
\label{eq:scalarsol}
\rho^{(\textrm{cl})}_k \left( \tau \right) & = & - i \frac{\sqrt{\pi}}{2\sqrt{2}} \frac{H}{M_{\textrm{Pl}} \sqrt{Q} k} \left(- \tau \right)^{\frac{5}{2}} H^{(1)}_{\frac{5}{2}} \left( - c_L k \tau \right), \\
\label{eq:scalarscal}
\delta \varphi^{(\textrm{cl})}_k \left( \tau \right) & = & - \frac{\sqrt{\pi}}{2} H \left( - \tau \right)^{\frac{3}{2}} H^{(1)}_{\frac{3}{2}} \left( - k \tau \right), \\
\label{eq:vector}
\pi^{(\textrm{cl})}_{\perp P k} \left( \tau \right) & = & i \frac{\sqrt{\pi}}{2\sqrt{2}} \frac{H}{M_{\textrm{Pl}} \sqrt{Q}} \left( - \tau \right)^{\frac{5}{2}} H^{(1)}_{\frac{5}{2}} \left( - c_T k \tau \right), \\
\label{eq:tensor}
\gamma^{\lambda (\textrm{cl})}_k \left( \tau \right) & = & \sqrt{\frac{\pi}{2}} \frac{H}{M_{\textrm{Pl}}} \left( - \tau \right)^{\frac{3}{2}} H^{(1)}_{\frac{3}{2}} \left( - k \tau \right),
\end{eqnarray}
where $H^{(1)}_{\nu}$ denote Hankel functions of the first kind. Modes with Hankel function of order $3/2$ are typical for the single-field model, while modes with Hankel function of order $5/2$ are typical for models with broken spatial diffeomorphism symmetry. The perturbation of scalar field added to the solid matter in the model studied in this paper is associated with order of Hankel function $3/2$, as well as perturbation of field breaking time diffeomorphism symmetry included in inflationary model studied in \cite{rew}. Note that the vector-to-scalar ratio for the solid matter perturbations is defined as $\left|\pi^{(\textrm{cl})}_{\perp}/(k \rho^{(\textrm{cl})})\right|^2$, and in the long wavelength limit it is
\begin{eqnarray}
s = \lim\limits_{k \tau \to 0^{-}} \left| \frac{ \pi^{(\textrm{cl})}_{\perp k} }{ k \rho^{(\textrm{cl})}_k }\right|^2
= \left(\frac{c_L}{c_T}\right)^5,
\end{eqnarray}
the same as for the solid inflation model \cite{endlich}. Because of the constraint on sound speeds (\ref{eq:soundconstraint}), this quantity is smaller than $\left( 1 / 3 \right)^{5/2} ( 1 + \mathcal{O} ( \epsilon) ) \approx 0.064$. 

\vskip 1mm
The classical modes are needed for calculating correlation functions for perturbations. It is conventional to calculate the scalar correlation functions for quantity $\zeta$ which parametrizes the curvature perturbation and is conserved in the superhorizon limit, so that it preserves the information about primordial perturbations generated during inflation even in the reheating era. However, as pointed out by authors of \cite{endlich}, in solid inflation model, and consequently also in our combined model, there is a mild time-dependence of $\zeta$ during inflation. For now, by restricting ourselves to the leading order of slow-roll approximation we disregard this time-dependence, and we will discuss it in subsection \ref{0000}. Following the definition of $\zeta$ in \cite{weinberg} we can express it in terms of variables used in this paper as
\begin{eqnarray}
\label{eq:zeta}
\zeta_k = \frac{1}{3 \epsilon} \left[ \frac{\pm \sqrt{p}}{\sqrt{2} M_{\textrm{Pl}} H} \left( \dot{\delta \varphi}_k - 3 H \delta \varphi_k \right) - Q k^2 \rho_k \right],
\end{eqnarray}
where only the leading order terms in slow-roll approximation have been kept.

\vskip 1mm
Two and three-point correlation functions define spectra and bispectra for every perturbation. In this section we need power spectrum for scalar and tensor perturbations and scalar bispectrum, and for them we use definitions
\begin{eqnarray}
\left<\zeta_{\mathbf{k}_1}\zeta_{\mathbf{k}_2}\right> & = & \left( 2 \pi \right)^3 \delta^{(3)} \left( \mathbf{k}_1 + \mathbf{k}_2 \right) \mathcal{P}_{\zeta} \left( k_1 \right), \\
\left<\gamma_{\mathbf{k}_1 ij}\gamma_{\mathbf{k}_2 lm}\right> & = & \left( 2 \pi \right)^3 \delta^{(3)} \left( \mathbf{k}_1 + \mathbf{k}_2 \right) \Pi_{ijlm} \left( \mathbf{k}_1 \right) \mathcal{P}_{\gamma} \left( k_1 \right), \\
\left<\zeta_{\mathbf{k}_1}\zeta_{\mathbf{k}_2}\zeta_{\mathbf{k}_3}\right> & = & \left( 2 \pi \right)^3 \delta^{(3)} \left( \mathbf{k}_1 + \mathbf{k}_2 + \mathbf{k}_3 \right) \mathcal{B}_{\zeta} \left( \mathbf{k}_1, \mathbf{k}_2, \mathbf{k}_3 \right),
\end{eqnarray}
where $\Pi_{ijlm}(\mathbf{k}) = \sum_{\lambda} e^{\lambda}_{\mathbf{k} ij} e^{\lambda *}_{\mathbf{k} lm}$, and all perturbations are evaluated in the late time limit when inflation ends, $\tau_{\textrm{e}} \to 0^{-}$. For scalar and tensor power spectrum we obtain
\begin{eqnarray}
\label{eq:powerspectra}
\mathcal{P}_\zeta(k)=\frac{H^2}{4M^2_{\textrm{Pl}}\epsilon^2}\left(p+\frac{Q}{c^5_L}\right)\frac{1}{k^3}, \quad \mathcal{P}_\gamma(k)=\frac{H^2}{M^2_{\textrm{Pl}}}\frac{1}{k^3},
\end{eqnarray}
with the tensor-to-scalar ratio
\begin{eqnarray}
\label{eq:tts}
r = \frac{\mathcal{P}_{\gamma}}{\mathcal{P}_{\zeta}} = \frac{4 c_L^5 \epsilon^2}{\epsilon + \left( c_L^5 - 1 \right) p}.
\end{eqnarray}
The CMB observations lead to the restriction $r < 0.064$, \cite{planck18, bicep}.

\subsection{Primordial non-Gaussianity}

Three-point correlation functions measure non-Gaussianity caused by nonlinear effects. They can be calculated with the use of in-in formalism \cite{weinberginin} with the dominant contribution to them given by
\begin{eqnarray}
\label{eq:inin}
\left< O \left( \tau \right) \right> = - i \int\limits_{- \infty}^{\tau} a \left( \tau^\prime \right) d \tau^\prime \left< 0 \right| \left[ O \left( \tau \right), H_{\textrm{int}} \left( \tau^\prime \right) \right] \left| 0 \right>,
\end{eqnarray}
with $O$ being product of three perturbations for which the three-point correlation function is calculated. The scalar three-point correlation function corresponds to $O ( \tau ) = \zeta_{\mathbf{k}_1} ( \tau ) \zeta_{\mathbf{k}_2} ( \tau )$ $\zeta_{\mathbf{k}_3} ( \tau )$ with the interaction Hamiltonian given by the scalar cubic action, which in the leading order of the slow-roll approximation is of the form
\begin{eqnarray}
S^{(3)}_{\textrm{S}} = \int d^4 x a^3 \bigg[ - \frac{8}{81} \left( \frac{2}{3} F_Y + \widetilde{F} \right) \left( \rho_{, i i} \right)^3 + \frac{8}{27} \left( F_Y + \widetilde{F} \right) \rho_{, i i} \rho_{, j k} \rho_{, j k} - \\
- \frac{8}{27} F_Y \rho_{, i j} \rho_{, i k} \rho_{, j k} \pm \frac{4 \sqrt{2}}{9 M_{\textrm{Pl}}} \frac{\widetilde{F}}{\sqrt{p}} \left( \rho_{, i j} \rho_{, i j} - \frac{1}{3} \left( \rho_{, i i} \right)^2 \right) \delta \varphi \bigg], \nonumber
\end{eqnarray}
where $\widetilde{F}$ is defined by
\begin{eqnarray}
\label{eq:tildeF}
\widetilde{F} = \pm \frac{M_{\textrm{Pl}}}{\sqrt{2}} \sqrt{p} \left( F_{\varphi Y} + F_{\varphi Z} \right) = X \left( F_{X Y} + F_{X Z} \right).
\end{eqnarray}
The last equality is the consequence of the slow-roll approximation and it holds in the leading order of it, see appendix \ref{app:a}. Finally, the dominant contribution to the scalar bispectrum is given by
\begin{eqnarray}
\label{eq:scalarbispectrum}
\mathcal{B}_{\zeta} \left( \mathbf{k}_1, \mathbf{k}_2, \mathbf{k}_3 \right) = \frac{H^2}{2 M_{\textrm{Pl}}^6 c_L^{12} \epsilon^3} \frac{1}{\left( k_1 k_2 k_3 \right)^3} \bigg[ \widetilde{Q} \left( \mathbf{k}_1, \mathbf{k}_2, \mathbf{k}_3 \right) \Lambda \left( k_1, k_2, k_3 \right) + \\
+ c_L^2 \widetilde{F} \left( Q^{(2,3)} \left( \mathbf{k}_1, \mathbf{k}_2, \mathbf{k}_3 \right) \Omega \left( k_1, c_L k_2, c_L k_3 \right) + 2 \textrm{ permutations} \right) \nonumber \bigg],
\end{eqnarray}
where
\begin{eqnarray}
& & \widetilde{Q} \left( \mathbf{k}_1, \mathbf{k}_2, \mathbf{k}_3 \right) = - \left( F_Y + \widetilde{F} \right) \frac{k_1^2 \left( \mathbf{k}_2 \cdot \mathbf{k}_3 \right)^2 + 2 \textrm{ permutations}}{\left(k_1 k_2 k_3\right)^2} + \\
& & + 3 F_Y \frac{\left( \mathbf{k}_1 \cdot \mathbf{k}_2 \right)\left( \mathbf{k}_1 \cdot \mathbf{k}_3 \right)\left( \mathbf{k}_2 \cdot \mathbf{k}_3 \right)}{\left( k_1 k_2 k_3 \right)^2} + \frac{2}{3} F_Y + \widetilde{F}, \nonumber\\
& & Q^{(A,B)} \left( \mathbf{k}_1, \mathbf{k}_2, \mathbf{k}_3 \right) = \frac{1}{6} - \frac{1}{2} \frac{\left(\mathbf{k}_A \cdot \mathbf{k}_B \right)^2}{\left(k_A k_B\right)^2}, \qquad A, B = 1, 2, 3, \\
& & \Lambda \left( k_1, k_2, k_3 \right) = - \frac{1}{27 \left( \sum\limits_i k_i \right)^3} \Bigg[ 3 \sum\limits_i k_i^6 + 9 \sum\limits_{i \neq j} k_i^5 k_j + 12 \sum\limits_{i \neq j} k_i^4 k_j^2 + \\
& & + 6 \sum\limits_{i \neq j} k_i^3 k_j^3 + 18 \left( \prod\limits_i k_i \right) \sum\limits_{i \neq j} k_i^2 k_j + 18 \left( \prod\limits_i k_i \right) \sum\limits_i k_i^3 + 20 \left( \prod\limits_i k_i \right)^2 \nonumber \Bigg], \\
& & \Omega \left( \mathcal{A}, b, c \right) = \frac{1}{3} \left[ \gamma_{\textrm{EM}} - N_{\textrm{min}} + \ln \left( \frac{\mathcal{A} + b + c}{k_{\textrm{min}}} \right) + \mathcal{O} \left( \frac{\mathcal{A} + b + c}{e^{N_{\textrm{min}}} k_{\textrm{min}}}\right) \right] \mathcal{A}^3 - \\
& & - \frac{1}{9 \left( \mathcal{A} + b + c \right)^2} \big[ b^5 + 2 b^4 c + 2 b^3 c^2 + 2 b^2 c^3 + 2 b c^4 + c^5 + \nonumber\\
& & + 2 \mathcal{A} \left( b^4 + b^3 c + b^2 c^2 + b c^3 + c^4 \right) + 2 \mathcal{A}^2 \left( 2 b^3 + 3 b^2 c + 3 b c^2 + 2 c^3 \right) + \nonumber\\
& & + \mathcal{A}^3 \left( 10 b^2 + 17 b c + 10 c^2 \right) + 11 \mathcal{A}^4 \left( b + c \right) + 4 \mathcal{A}^5 \big], \nonumber
\end{eqnarray}
with $\gamma_{\textrm{EM}}$ denoting the Euler--Mascheroni constant, $N_{\textrm{min}}$ is the minimal number of $e$-folds, and $k_{\textrm{min}}$ is the comoving wavenumber corresponding to longest mode observable today. We have evaluated the integral (\ref{eq:inin}) with the upper limit corresponding to conformal time of the end of inflation, which can be rewritten as $\tau_{\textrm{e}} = - e^{-N_{\textrm{min}}} / k_{\textrm{min}}$, and with infinitesimally tilted contour, $\tau \to \tau (1 - i \varepsilon)$, with $\varepsilon \to 0^{+}$, which provides projection onto the right vacuum. Since the sound speed of solid matter scalar perturbation $c_L$ cannot exceed sound speed of scalar field perturbation, which equals the light speed, the horizon crossing is associated with the latter quantity. Therefore, the sound speed $c_L$ is not hidden in $k_{\textrm{min}}$ and it appears in the argument of logarithm in formula for $\Omega \left( \mathcal{A}, b, c \right)$. For the same reason, also the tensor-scalar-scalar bispectrum calculated in the following section will depend on $c_L$ in such way.

\vskip 1mm
Note that the minimal wavenumber of the longest mode of the observational relevance today $k_{\textrm{min}}$ is often defined only by its order of magnitude. Consequently instead of equality in expression for the time when inflation ends there is only relation of being of the same order, $\tau_{\textrm{e}} \sim - e^{-N_{\textrm{min}}} / k_{\textrm{min}}$. In this convention there is no reason to keep terms much smaller that $N_{\textrm{min}}$ when they are added to or subtracted from it, and reader may disregard them. In this paper we use the mentioned equality as a definition for the exact value of $k_{\textrm{min}}$, and therefore, we are keeping all such terms.

\vskip 1mm
Absolute value of the scalar bispectrum is largest in the squeezed limit, $k_1 \ll k_2 \approx k_3$, where unlike for standard local shape the bispectrum diverges to plus or minus infinity depending on direction in which the limit is approached. Encode this direction by angle $\theta$ defined as $\mathbf{k}_1 \cdot \mathbf{k}_2 = - k_1 k_2 \cos \theta$, i.e. the angle between two sides of a triangle with lengths $k_1$ and $k_2$, where the length of its third side is $k_3$. Using the standard definition of local nonlinearity parameter \cite{komatsu, maldacena},
\begin{eqnarray}
f_{\textrm{NL}}^{(\textrm{local})} = \frac{5}{12} \lim\limits_{k_1 \to 0} \frac{\mathcal{B}_{\zeta} \left( k_1, k_2, k_3 \right)}{\mathcal{P}_{\zeta} \left( k_1 \right) \mathcal{P}_{\zeta} \left( k_2 \right)},
\end{eqnarray}
we find
\begin{eqnarray}
\label{eq:local}
f_{\textrm{NL}}^{(\textrm{local})} = \frac{\epsilon}{\left[ \epsilon + \left( c_L^5 - 1 \right) p \right]^2} \Bigg\{ \frac{25}{27} \frac{1}{c_L^2} \left( 1 - 3 \cos^2 \theta \right) \frac{F_Y}{F} + \\
+ \left[ \frac{50}{9} \frac{1}{c_L^2} \cos^2 \theta - \frac{10}{9} \mathcal{N} \left( 1 - 3 \cos^2 \theta \right) \right] \frac{\widetilde{F}}{F} \nonumber \Bigg\},
\end{eqnarray}
where $\mathcal{N}$ is of the same order as minimal number of $e$-folds, more precisely
\begin{eqnarray}
\mathcal{N} = N_{\textrm{min}} - \gamma_{\textrm{EM}} + \frac{1}{3} \left( 1 + c_L \right) \left( c_L^2 - c_L + 4 \right),
\end{eqnarray}
so that $\mathcal{N}$ differs from $N_{\textrm{min}}$ at most by factor $8 / 3 - \gamma_{\textrm{EM}} \approx 2.09$. A similar dependence of the scalar bispectrum in the squeezed limit on angle $\theta$ appears not only in solid inflation \cite{endlich}, but also in inflationary model studied by authors of \cite{rew}. Observations of the CMB anisotropies are consistent with the theory if $f_{\textrm{NL}}^{(\textrm{local})} = 0.8 \pm 5$ \cite{planck15}, so that $F_Y / F$ and $\widetilde{F} / F$ must be suppressed by slow-roll parameters.

\subsection{Scalar spectral tilt}\label{0000}

So far we have restricted ourselves to only the leading order of the slow-roll approximation. However, to calculate the scalar spectral tilt, the next to leading order must be considered. This include time-dependence of the Hubble parameter and slow-roll parameters, $H = H_{\textrm{c}} \left( \tau / \tau_{\textrm{c}} \right)^{\epsilon_{\textrm{c}}}$, $\epsilon = \epsilon_{\textrm{c}} \left( \tau / \tau_{\textrm{c}} \right)^{- \eta_{\textrm{c}}}$, $p = p_{\textrm{c}} \left( \tau / \tau_{\textrm{c}} \right)^{- \eta_{p, \textrm{c}}}$, $Q = Q_{\textrm{c}} \left( \tau / \tau_{\textrm{c}} \right)^{- \eta_{Q, \textrm{c}}}$, $c_L = c_{L,\textrm{c}} \left( \tau / \tau_{\textrm{c}} \right)^{- \eta_{L,\textrm{c}}}$, and slow-roll correction to time-dependence of the scale factor $a = a_{\textrm{c}} \left( \tau / \tau_{\textrm{c}} \right)^{- 1 - \epsilon_{\textrm{c}}}$, with subscript $\textrm{c}$ indicating the time when the longest mode observable today exists the horizon, $- k_{\textrm{min}} \tau_{\textrm{c}} = 1$.

\vskip 1mm
First we consider the quadratic action (\ref{eq:reducedaction}) without terms with $\mathcal{C}_3$ and $\mathcal{C}_4$, and derive equations for scalar perturbations from it keeping terms up to the first order in the slow-roll parameters,
\begin{eqnarray}
\rho^{\prime \prime} + \frac{- 4 - 2 \epsilon_{\textrm{c}} - \eta_{Q,\textrm{c}}}{\tau} \rho^{\prime} + c_L^2 k^2 \rho & = & 0, \\
\delta \varphi^{\prime \prime} + \frac{- 2 - \epsilon_{\textrm{c}}}{\tau} \delta \varphi^{\prime} + k^2 \delta \varphi & = & 0.
\end{eqnarray}
By solving these equations and matching canonically normalized perturbations to the free wave function of the Minkowski space with cosmic time replaced by the conformal time in the limit of very early time, we find the classical modes
\begin{eqnarray}
\rho^{(\textrm{cl})}_k \left( \tau \right) & = & - i \frac{\sqrt{\pi}}{2\sqrt{2}} \frac{H_{\textrm{c}}}{M_{\textrm{Pl}} \sqrt{Q_{\textrm{c}}} k} \left( - \tau_{\textrm{c}} \right)^{ - \epsilon^{(Q)}_{\textrm{c}} } \left( 1 - 2 \epsilon_{\textrm{c}} \right) e^{i \frac{\pi}{2} \left( \epsilon^{(Q)}_{\textrm{c}} + \frac{5}{2} \eta_{L,\textrm{c}} \right)} \\
& & \left( - \tau \right)^{\frac{5}{2} + \epsilon^{(Q)}_{\textrm{c}} } H^{(1)}_{\frac{5}{2} + \epsilon^{(Q)}_{\textrm{c}} + \frac{5}{2} \eta_{L,\textrm{c}}} \left( - c_L \left( \tau \right) \left( 1 + \eta_{L,\textrm{c}} \right) k \tau \right), \nonumber\\
\delta \varphi^{(\textrm{cl})}_k \left( \tau \right) & = & - \frac{\sqrt{\pi}}{2} H_{\textrm{c}} \left( - \tau_{\textrm{c}} \right)^{\epsilon_{\textrm{c}}} \left( 1 - \epsilon_{\textrm{c}} \right) e^{i \epsilon_{\textrm{c}}} \left( - \tau \right)^{\frac{3}{2} + \epsilon_{\textrm{c}}} H^{(1)}_{\frac{3}{2} + \epsilon_{\textrm{c}}} \left( - k \tau \right),
\end{eqnarray}
where $\epsilon^{(Q)} = \epsilon + \frac{1}{2} \eta_{Q}$. Inserting these modes into (\ref{eq:zeta}) we can calculate the power spectrum of perturbation $\zeta$ evaluated at the time when inflation ends,
\begin{eqnarray}
\label{eq:power2}
\mathcal{P}_{\zeta} \left( k \right) & = & \frac{H_{\textrm{c}}^2}{4 M_{\textrm{Pl}}^2 \epsilon_{\textrm{c}}^2} \frac{1}{k^3} \left( \frac{\tau_{\textrm{e}}}{\tau_{\textrm{c}}} \right)^{2 \eta_{\textrm{c}} + 2 \epsilon_{\textrm{c}}} \left( - k \tau_{\textrm{e}} \right)^{- 2 \epsilon_{\textrm{c}}} \Biggl[ p_{\textrm{c}} \left( \frac{\tau_{\textrm{e}}}{\tau_{\textrm{c}}} \right)^{- \eta_{p,\textrm{c}}} + \\
& & + \frac{Q_{\textrm{c}}}{c_{L,\textrm{c}}^5} \left( \frac{\tau_{\textrm{e}}}{\tau_{\textrm{c}}} \right)^{- \eta_{Q,\textrm{c}} + 5 \eta_{L,\textrm{c}}} \left( - k \tau_{\textrm{e}} \right)^{- \eta_{Q,\textrm{c}} - 5 \eta_{L,\textrm{c}}} \Biggr]. \nonumber
\end{eqnarray}
Of course, because of considering only the part of the quadratic action (\ref{eq:reducedaction}) the scalar power spectrum calculated in this way is not correct.

\vskip 2mm
In order to calculate the scalar spectral tilt correctly, the omitted terms with $\mathcal{C}_3$ and $\mathcal{C}_4$ in the quadratic action (\ref{eq:reducedaction}) must be taken into account. Assuming their smallness, we can handle them as interaction terms,
\begin{eqnarray}
\label{eq:quadraticinteraction}
\delta H^{(2)}_{\textrm{int}} = \int \frac{d^3 k}{(2 \pi)^3} a^3 H^2 \frac{Q}{p}
\left( \pm 6 \sqrt{2} M_{\textrm{Pl}} \mathcal{C}_3 \sqrt{p} k^2 \textrm{Re}\left\{ \rho \delta \varphi^{*} \right\} - 9 \mathcal{C}_4 \left| \delta \varphi \right|^2 \right),
\end{eqnarray}
and employ in-in formalism to calculate correction to the two-point correlation function associated with the interaction Hamiltonian $\delta H^{(2)}_{\textrm{int}}$,
\begin{eqnarray}
\label{eq:inin2}
\delta \left< \zeta_{\mathbf{k}_1} (\tau) \zeta_{\mathbf{k}_2} (\tau) \right> = - i \int\limits_{-\infty}^{\tau} a (\tau^{\prime}) d \tau^{\prime} \left< 0 \right| \left[ \zeta_{\mathbf{k}_1} (\tau) \zeta_{\mathbf{k}_2} (\tau), \delta H^{(2)}_{\textrm{int}} (\tau^{\prime}) \right] \left| 0 \right>.
\end{eqnarray}
Inserting classical modes (\ref{eq:scalarsol}) and (\ref{eq:scalarscal}) into these formulas, using expression for perturbation $\zeta$ (\ref{eq:zeta}), taking into account the time-dependence of slow-roll parameters, evaluating the integral with $\varepsilon$-tilted contour, taking the superhorizon limit, $\left| k \tau \right| \ll 1$, and adding the result to the scalar power spectrum (\ref{eq:power2}) we obtain
\begin{eqnarray}
\label{eq:scalarpowerspectrum}
\mathcal{P}_{\zeta} \left( k \right) = \frac{H^2 Q}{M_{\textrm{Pl}}^2 \epsilon^2} \frac{1}{k^3} \left[ \lambda_0 + \delta \lambda_0 + \delta \lambda_1 \ln \left( - k \tau \right) \right],
\end{eqnarray}
where
\begin{eqnarray}
\lambda_0 & = & \frac{1}{4} \left[ \frac{p}{Q} \left( - k \tau \right)^{- 2 \epsilon_{\textrm{c}} } + \frac{1}{c_L^5} \left( - k \tau \right)^{- 2 \tilde{\epsilon}_{\textrm{c}} } \right], \\
\delta \lambda_0 & = & - \frac{1}{2} \frac{\mathcal{C}_3}{c_L^5} \left( \frac{4}{3} - \gamma_{\textrm{EM}} + \frac{1}{3} c_L \left( 3 + c_L^2 \right) - \ln \left( 1 + c_L \right) \right) \left( - k \tau \right)^{ \frac{1}{2} \tilde{\eta}_{3,\textrm{c}}} - \\
& & - 6 \mathcal{C}_4 \left( \frac{4}{3} - \gamma_{\textrm{EM}} + \frac{3}{8} \pi - \ln 2 \right) \left( - k \tau \right)^{ \frac{7}{8} \tilde{\eta}_{4,\textrm{c}}}, \nonumber\\
\delta \lambda_1 & = & \frac{1}{2} \frac{\mathcal{C}_3}{c_L^5} \left( - k \tau \right)^{ \frac{1}{2} \tilde{\eta}_{3,\textrm{c}}} + 6 \mathcal{C}_4 \left( - k \tau \right)^{ \frac{7}{8} \tilde{\eta}_{4,\textrm{c}}},
\end{eqnarray}
with $\tilde{\epsilon} = \epsilon + \eta_Q / 2 + 5 \eta_L / 2$, $\tilde{\eta}_3 = \eta_3 + \eta_Q / 2 - \eta_p / 2 - 5 \eta_L / 2$, $\tilde{\eta}_4 = \eta_4 + \eta_Q - \eta_p$, and in addition to slow-roll parameters defined so far $\eta_3 = \dot{\mathcal{C}}_3 / \left( H \mathcal{C}_3 \right)$ and $\eta_4 = \dot{\mathcal{C}}_4 / \left( H \mathcal{C}_4 \right)$.

\vskip 1mm
There are two kinds of time-dependence of the scalar power spectrum in the superhorizon limit (\ref{eq:scalarpowerspectrum}). One of them arises from mild time-dependence of slow-roll parameters and the second type is due to logarithm in formula (\ref{eq:scalarpowerspectrum}). Since $\mathcal{P}_{\zeta} = \left| \zeta \right|^2$, the second kind of time-dependence of the power spectrum is a consequence of the fact that the curvature perturbation $\zeta$ is not conserved after its exits the horizon, unlike in standard inflationary models. This is a feature of also the solid inflation model as well as other models with broken diffeomorphism symmetry. Due to this time-dependence, the primordial power spectrum of perturbations after the end of inflation cannot be obtained by evaluating the power spectrum (\ref{eq:scalarpowerspectrum}) at the time of horizon crossing, as it can be often done with other inflationary models, but by evaluating it at the end if inflation, as we have already done with (\ref{eq:power2}). The scalar spectral tilt is then
\begin{eqnarray}
\label{eq:nsm1}
& & n_{\textrm{s}} - 1 = \Biggl. \frac{d \ln \left( k^3 \mathcal{P}_{\zeta} \left( k \right) \right)}{d \ln k} \Biggr|_{ k = k_{\textrm{min}}, \tau = \tau_{\textrm{e}} } \approx \\
& \approx & \Biggl. \Biggl[ - \frac{1}{2} \frac{p}{Q} e^{2 N_{\textrm{min}} \epsilon_{\textrm{c}}} \epsilon_{\textrm{c}} - \frac{1}{2} \frac{1}{c_{L,\textrm{c}}^5} e^{2 N_{\textrm{min}} \tilde{\epsilon}_{\textrm{c}}} \tilde{\epsilon}_{\textrm{c}} + \nonumber\\
& & + \frac{1}{2} \frac{\mathcal{C}_3}{c_L^5} e^{- \frac{1}{2} N_{\textrm{min}} \tilde{\eta}_{3,\textrm{c}} } \left( 1 + \frac{1}{2} N_{\textrm{min}} \tilde{\eta}_{3,\textrm{c}} \right) + 6 \mathcal{C}_4 e^{- \frac{7}{8} N_{\textrm{min}} \tilde{\eta}_{4,\textrm{c}} } \left( 1 + \frac{7}{8} N_{\textrm{min}} \tilde{\eta}_{4,\textrm{c}} \right) \Biggr] \nonumber\\
& & \Biggl[ \frac{1}{4} \frac{p}{Q} e^{2 N_{\textrm{min}} \epsilon_{\textrm{c}}} + \frac{1}{4} \frac{1}{c_L^5} e^{2 N_{\textrm{min}} \tilde{\epsilon}_{\textrm{c}}} + \nonumber\\
& & + \frac{1}{2} \frac{\mathcal{C}_3}{c_L^5} e^{- \frac{1}{2} N_{\textrm{min}} \tilde{\eta}_{3,\textrm{c}}} N_{\textrm{min}} + 6 \mathcal{C}_4 e^{- \frac{7}{8} N_{\textrm{min}} \tilde{\eta}_{4,\textrm{c}}} N_{\textrm{min}} \Biggr]^{-1} \Biggr|_{ \tau = \tau_{\textrm{e}} }, \nonumber
\end{eqnarray}
where the approximative equality corresponds to keeping only terms contributing up to the first order of the slow-roll approximation. This is also consistent with form of the scalar quadratic action (\ref{eq:reducedaction}). We can also see that coefficient of the late time-dependence of $\zeta$ in the superhorizon limit $m_{\zeta}$ is related to the scalar power spectrum $n_{\textrm{s}}$ in the long wavelength limit,
\begin{eqnarray}
m_{\zeta} = \Biggl. \frac{d \ln \zeta}{d \ln \tau} \Biggr|_{\tau = \tau_{\textrm{e}}, k = k_{\textrm{min}}} = \Biggl. \frac{d}{d \ln k} \ln \left( \sqrt{ k^3 \mathcal{P}_{\zeta} \left( k \right) } \right) \Biggr|_{ k = k_{\textrm{min}}, \tau = \tau_{\textrm{e}} } = \frac{n_{\textrm{s}} - 1}{2}.
\end{eqnarray}
Therefore, in order to satisfy observatory restriction on scalar spectral index the time-dependence of perturbation $\zeta$ after exiting the horizon cannot be set to be negligible by an appropriate choice of parameters of the theory. Our combined model shares this feature with the solid inflation model, see equation (6.30) in \cite{endlich}.

\begin{figure}[!htb]
\centering
\sbox0{
\includegraphics[scale=0.32]{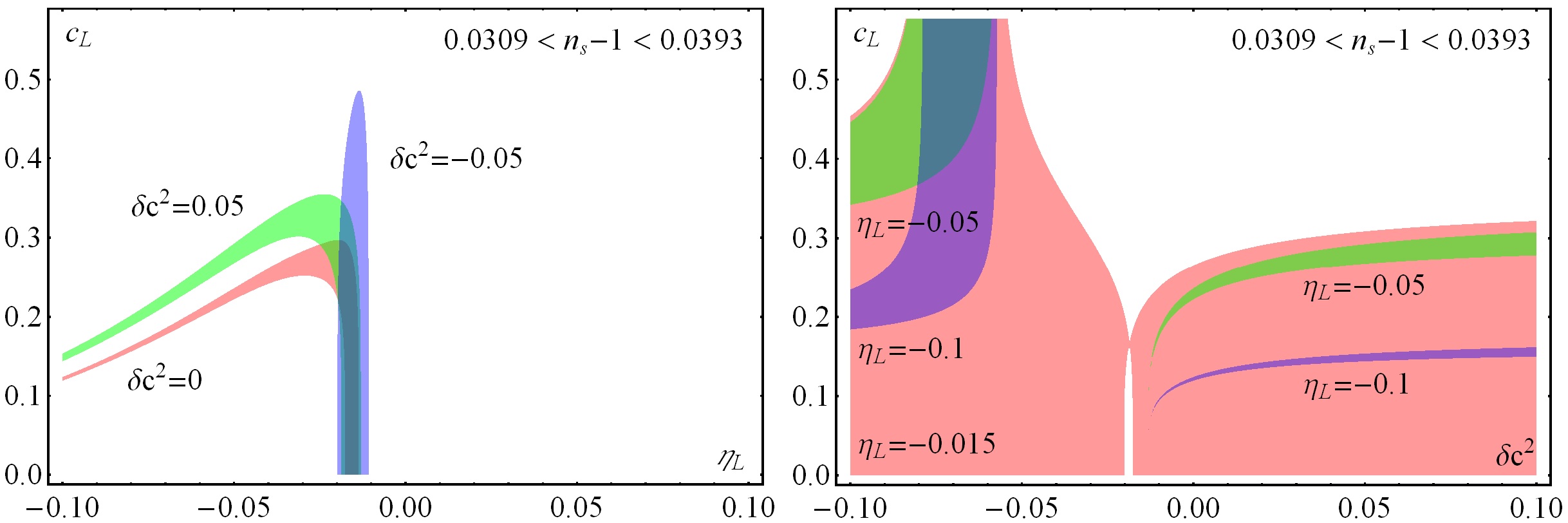}
}
%\centering
\begin{minipage}{\wd0}
\usebox0
\linespread{1}
\setlength{\abovecaptionskip}{-8pt plus 0pt minus 0pt}
\caption{{\footnotesize Regions in parameter space of the model consistent with the observed scalar spectral index. Left panel: $c_L$-$\eta_L$-plane with $p = q = 10^{-2}$ and with other slow-roll parameters set to zero except for $\delta c^2 = 3 c_T^2 / 4 - c_L^2 - 1$ set to $\delta c^2 = 0$ (red area), $\delta c^2 = 0.05$ (green area), and $\delta c^2 = -0.05$ (blue area). Right panel: $c_L$-$\delta c^2$-plane with the same choice of fixed slow-roll parameters for $\eta_L = -0.015$ (red area), $\eta_L = -0.05$ (green area), and $\eta_L = -0.1$ (blue area).}}
\label{fig:00}
\end{minipage}
\end{figure}

The value of scalar spectral tilt given by CMB observations is $n_{\textrm{s}} = 0.9649 \pm 0.0042$ \cite{planck18}. As shown in Fig. \ref{fig:00} the parameter space of the model predicting the spectral tilt (\ref{eq:nsm1}) accommodates regions consistent with this restriction. These regions are depicted with the use of (\ref{eq:nsm1}) for both $p$ and $q$ set to the value $10^{-2}$ and with other slow-roll parameters set to zero except for $\eta_L$ and $\delta c^2$ defined based on relation between two sound speeds (\ref{eq:soundconstraint}) as $3 c_T^2 / 4 - c_L^2 = 1 + \delta c^2$.

\vskip 1mm
Integral (\ref{eq:inin}) as well as its special form (\ref{eq:inin2}) represent only the leading contribution to correlation functions. The simplest next to leading contribution to the scalar two-point correlation function is given by
\begin{eqnarray}
\label{eq:inin22}
\delta^{(2)} \left< \zeta_{\mathbf{k}_1} (\tau) \zeta_{\mathbf{k}_2} (\tau) \right> & = & - \int\limits_{-\infty}^{\tau} a (\tau_1) d \tau_1 \int\limits_{-\infty}^{\tau_1} a (\tau_2) d \tau_2 \\
& & \left< 0 \right| \left[ \left[ \zeta_{\mathbf{k}_1} (\tau) \zeta_{\mathbf{k}_2} (\tau), \delta H^{(2)}_{\textrm{int}}(\tau_1) \right], \delta H^{(2)}_{\textrm{int}}(\tau_2) \right] \left| 0 \right>. \nonumber
\end{eqnarray}
Replacing lower limits of integrals in this formula by the time when the scalar field perturbation $\delta \varphi$ with wavenumber $k$ exits the horizon, $- \infty \to - 1/k$, i.e. disregarding phase when both scalar perturbations oscillate, we can evaluate this contribution to the late time two-point correlation function,
\begin{eqnarray}
\delta^{(2)} \left< \zeta_{\mathbf{k}_1} (\tau) \zeta_{\mathbf{k}_2} (\tau) \right> & = & \left( 2 \pi \right)^3 \delta^{(3)} \left( \mathbf{k}_1 + \mathbf{k}_2 \right) \frac{H^2}{M_{\textrm{Pl}}^2} \frac{Q^2}{\epsilon^2 p} \frac{1}{k^3} \Biggl[ \frac{9}{50} \frac{\mathcal{C}_3^2}{c_L^5} + \\
& & + \left( \frac{12}{25} + \frac{2}{c_L^5} \right) \mathcal{C}_3 \mathcal{C}_4 + 2 \mathcal{C}_4^2 + 6 \left( \frac{\mathcal{C}_3}{c_L^5} + \mathcal{C}_4 \right) \mathcal{C}_4 \ln \left( - k \tau \right) + \nonumber\\
& & + 9 \left( \frac{p}{Q} \frac{\mathcal{C}_3^2}{c_L^5} + \frac{\mathcal{C}_3 \mathcal{C}_4}{c_L^5} + 2 \mathcal{C}_4^2 \right) \left( \ln \left( -k \tau \right) \right)^2 \Biggr]. \nonumber
\end{eqnarray}
This correction to the scalar two-point function quadratic in parameters $\mathcal{C}_3$ and $\mathcal{C}_4$ is not the only one correction of the form (\ref{eq:inin22}). Actually, there is contribution not only from the interaction Hamiltonian (\ref{eq:quadraticinteraction}) but also contributions from cubic and higher order interactions corresponding to loop corrections.

\section{Clustering fossils}\label{sec:3}

Superhorizon tensor perturbation mode with polarization $\lambda$ and comoving wavevector $\mathbf{K}$ affects scalar perturbations causing quadrupole anisotropy in their power spectrum. Since in linear order of the perturbation theory tensor perturbations are decoupled from scalar perturbations, this anisotropy is a result of nonlinear effects measured by tensor-scalar-scalar three-point correlation function. The effect on the scalar power spectrum is of the form
\begin{eqnarray}
\label{eq:toquad}
\mathcal{P}_\zeta(k) \to \mathcal{P}_\zeta(\mathbf{k}) = \mathcal{P}_\zeta(k) \left( 1 + Q^{\lambda}_{ij}(K,k)\frac{k^ik^j}{k^2} \right).
\end{eqnarray}
Note that such change in scalar power spectrum does not affect the angular power spectrum coefficients $C_l$, which are usually used for describing statistical features of the CMB anisotropies, because they are defined as averaging which preserves only the trace of $Q_{ij}$, see appendix \ref{app:x}, and in this particular case its trace is zero due to its proportionality to the tensor perturbation mode. The quadrupole can be expressed through tensor-scalar-scalar bispectrum $\mathcal{B}_{\gamma\zeta\zeta}(k_1,k_2,k_3)$ in the squeezed limit with long-wavelength tensor mode defined as
\begin{eqnarray}
\label{eq:tssbispectrum}
\left< \gamma_{\mathbf{K}ij} \zeta_{\mathbf{k}_1} \zeta_{\mathbf{k}_2} \right> = \left( 2 \pi \right)^3 \delta^{(3)}\left( \mathbf{K} + \mathbf{k}_1 + \mathbf{k}_2 \right) \Pi_{ijlm} \left( \mathbf{K} \right) \frac{k_1^lk_2^m}{k_1k_2} \mathcal{B}_{\gamma\zeta\zeta}\left( K, k_1, k_2 \right),
\end{eqnarray}
where $K \ll k_1 \approx k_2$. Specifically, the quadrupole can be written as
\begin{eqnarray}
\label{eq:quadrupole}
Q^{\lambda}_{ij}(K,k) = \gamma^{\lambda}_{\mathbf{K}ij} \frac{\mathcal{B}_{\gamma\zeta\zeta}(K,k,k)}{\mathcal{P}_\gamma(K) \mathcal{P}_\zeta(k)}.
\end{eqnarray}
This effect is called clustering fossils and represents an alternative way of detecting primordial tensor perturbations in addition to attempts to measure them directly \cite{bicep2}.

\subsection{Bispectrum with long-wavelength tensor mode}

Our goal is to calculate the tensor-scalar-scalar bispectrum in the squeezed limit with superhorizon tensor perturbation. We employ in-in formalism by inserting $O \left( \tau \right) = \gamma_{\mathbf{K}ij} \left( \tau \right) \zeta_{\mathbf{k}_1} \left( \tau \right)$ $\zeta_{\mathbf{k}_2} \left( \tau \right)$ into the integral (\ref{eq:inin}), and evaluate it with time when the inflation ends $\tau_{\textrm{e}}$ as its upper limit. Part of the interaction Hamiltonian which contributes to this integral is given by the tensor-scalar-scalar cubic action. Omitting terms with negligible contribution to the squeezed bispectrum, it can be written as
\begin{eqnarray}
\label{eq:tensorscalarscalar}
S^{(3)}_{\gamma\zeta\zeta} = \int d^4 x a^3 \gamma_{i j} \bigg( \frac{1}{2} a^{-2} \delta \varphi_{, i} \delta \varphi_{, j} + A \rho_{, i k} \rho_{, j k} + B \rho_{, i j} \rho_{, k k} + \\
+ a^2 C \dot{\rho}_{, i} \dot{\rho}_{, j} + a^2 H q D \rho_{, i j} \dot{\rho} \pm \frac{\sqrt{p}}{\sqrt{2} M_{\textrm{Pl}}} E \rho_{, i j} \delta \varphi
\bigg), \nonumber
\end{eqnarray}
where
\begin{eqnarray}
A & = & - \frac{1}{3} X F_X - \frac{10}{9} F_Y - \frac{14}{9} F_Z, \\
B & = & - \frac{4}{9} X^2 F_{X X} + \frac{8}{9} F_Y + \frac{32}{27} F_Z - \frac{8}{9} \left( F_{X Y} + F_{X Z} \right), \nonumber\\
C & = & \frac{2}{9} \left( F_Y + F_Z \right), \nonumber\\
D & = & - \frac{2}{3} X F_X - \frac{4}{9} \left( F_Y + F_Z \right), \nonumber\\
E & = & \frac{\pm \sqrt{2} M_{\textrm{Pl}}}{\sqrt{p}} \left[ - \frac{2}{3} X F_{\varphi X} - \frac{4}{9} \left( F_{\varphi Y} + F_{\varphi Z} \right) \right] + D. \nonumber
\end{eqnarray}
Terms with $A$, $B$ and $C$ follow from direct expansion of function $F ( \varphi, X, Y, Z )$ in terms of solid matter perturbations, term with $D$ arises from interaction of solid matter with gravity through expansion of $\sqrt{-g}$, and term with $E$ is included because of direct interaction between scalar field and solid matter. Contributions from the Einstein--Hilbert part of the action are suppressed by slow-roll parameters, and therefore they are not included in (\ref{eq:tensorscalarscalar}). Comparing partial derivatives of the function $F$ with its perturbative expansion (\ref{eq:Fexpand}), which follows from the slow-roll restriction, we can rewrite these parameters as
\begin{eqnarray}
& & \frac{A}{F} = \frac{7 \mathcal{C}_2 - 1}{3} q  + \frac{4}{9} \frac{F_Y}{F}, \qquad \frac{B}{F} = \frac{8}{9} \left( \mathcal{C}_1 - 2 \mathcal{C}_2 \right) q - \frac{8}{27} \left( \frac{F_Y}{F} + 3 \frac{\widetilde{F}}{F} \right), \\
& & \frac{C}{F} = - \frac{1}{3} \mathcal{C}_2 q, \qquad \frac{D}{F} = \frac{2}{3} \left( \mathcal{C}_2 - 1 \right) q, \qquad \frac{E}{F} = \frac{1}{p} \left( 4 \mathcal{C}_3 q - \frac{4}{9} \frac{\widehat{F}}{F} \right) + \frac{D}{F}, \nonumber
\end{eqnarray}
where $\widehat{F} = \pm \sqrt{2p} M_{\textrm{Pl}}^{-1} \left( F_{\varphi Y} + F_{\varphi Z} \right)$ and according to (\ref{eq:tildeF}) it can be replaced by $2 \widetilde{F}$ in the leading order of the slow-roll approximation.

\vskip 1mm
Using modes (\ref{eq:scalarsol}), (\ref{eq:scalarscal}) and (\ref{eq:tensor}) within the in-in formalism for calculation of the tensor-scalar-scalar three-point function evaluated at the end of inflation, $\tau = \tau_{\textrm{e}}$, and keeping only terms which will dominate after taking the squeezed limit with long-wavelength tensor mode $\gamma_{\mathbf{K}ij}$, we find
\begin{eqnarray}
\label{eq:tss3point}
& & \left< \gamma_{\mathbf{K} ij} \left( \tau_{\textrm{e}} \right) \zeta_{\mathbf{k}_1} \left( \tau_{\textrm{e}} \right) \zeta_{\mathbf{k}_1} \left( \tau_{\textrm{e}} \right) \right> \stackrel{K \to 0}{\longrightarrow} \left( 2 \pi \right)^3 \delta^{(3)} \left( \mathbf{K} + \mathbf{k}_1 + \mathbf{k}_2 \right)  \frac{ H^4 }{ 8 M_{\textrm{Pl}}^4 \epsilon^2 } \\
& & \frac{ \Pi_{ijlm} \left( \mathbf{K} \right) }{ \left( K k_1 k_2 \right)^3 } \Bigg[ \left( \mathbf{k}_1 \right)_l \left( \mathbf{k}_2 \right)_m \left( p \mathcal{I} \left( K, k_1, k_2 \right) - \frac{3}{c_L^6} \frac{C}{F} \mathcal{I} \left( K, c_L k_1, c_L k_2 \right) \right) + \nonumber\\
& & + \frac{3}{c_L^{10}} \left( \frac{A}{F} \frac{ \left( \mathbf{k}_1 \cdot \mathbf{k}_2 \right) \left( \mathbf{k}_1 \right)_l \left( \mathbf{k}_2 \right)_m }{ \left( k_1 k_2 \right)^2 } + \frac{B}{F} \frac{ \left( \mathbf{k}_1 \right)_l \left( \mathbf{k}_1 \right)_m }{ k_1^2 } \right) \mathcal{J} \left( K, c_L k_1, c_L k_2 \right) - \nonumber\\
& & +\frac{ \left( \mathbf{k}_1 \right)_l \left( \mathbf{k}_1 \right)_m}{k_1^2} \left( \frac{3 q}{c_L^8} \frac{D}{F} \mathcal{Y} \left( K, c_L k_1, c_L k_2 \right) - \frac{6 p}{c_L^5} \frac{E}{F} \mathcal{Y} \left( K, c_L k_1, k_2 \right) \right) + \nonumber\\
& & + \textrm{ exchange} \left( 1 \leftrightarrow 2 \right) \Bigg], \nonumber
\end{eqnarray}
where integrals
\begin{eqnarray}
\mathcal{I} \left( a, b, c \right) & = & \textrm{Re} \left\{ \int\limits_{-\tau_{\textrm{e}}}^{\infty (1 - i \varepsilon)} \frac{d x}{x^2} \mathcal{U} (a x) \mathcal{U} (b x) \mathcal{U} (c x) e^{-i (a + b + c) x} \right\}, \\
\mathcal{J} \left( a, b, c \right) & = & \textrm{Re} \left\{ \int\limits_{-\tau_{\textrm{e}}}^{\infty (1 - i \varepsilon)} \frac{d x}{x^4} \mathcal{U} (a x) \mathcal{V} (b x) \mathcal{V} (c x) e^{-i (a + b + c) x} \right\}, \\
\mathcal{Y} \left( a, b, c \right) & = & \textrm{Re} \left\{ \int\limits_{-\tau_{\textrm{e}}}^{\infty (1 - i \varepsilon)} \frac{d x}{x^4} \mathcal{U} (a x) \mathcal{V} (b x) \mathcal{U} (c x) e^{-i (a + b + c) x} \right\}, 
\end{eqnarray}
are defined by polynomials $\mathcal{U} (x) = i - x$ and $\mathcal{V} (x) = 3 i - 3 x - i x^2$, which appear in expressions for Hankel functions of the first kind of the order $3/2$ and $5/2$, respectively. Integrals $\mathcal{J}$ and $\mathcal{Y}$ diverge logarithmically in the limit $\tau_{\textrm{e}} \to 0^{-}$, however, in the squeezed limit the logarithmic divergence of integral $\mathcal{J}$ is suppressed. By calculating these integrals in the considered limits we obtain
\begin{eqnarray}
\mathcal{I} \left( a, b, c \right) & \to & - \frac{ b^3 + 2 b^2 c + 2 b c^2 + c^3 }{ \left( b + c \right)^2 }, \\
\mathcal{J} \left( a, b, c \right) & \to & - \frac{ b^4 + b^3 c + b^2 c^2 + b c^3 + c^4 }{ b + c }, \\
\mathcal{Y} \left( a, b, c \right) & \to & c^3 \ln \left[ - \tau_{\textrm{e}} \left( b + c \right) \right] - \frac{1}{3} b^3 - b c^2 + c^3 \left( \gamma_{\textrm{EM}} - \frac{4}{3} \right).
\end{eqnarray}
After inserting these results into (\ref{eq:tss3point}), using the definition (\ref{eq:tssbispectrum}) and the same conventions as we have used in the previous section for calculation of the scalar bispectrum, we can express the dominant contribution to the tensor-scalar-scalar bispectrum in the squeezed limit as
\begin{eqnarray}
\label{eq:tssB}
\mathcal{B}_{\gamma\zeta\zeta} \left( K, k, k \right) = - \frac{3 H^4}{8 M_{\textrm{Pl}}^4 \epsilon^2} \frac{1}{ K^3 k^3 } \left[ \nu + \omega \ln \left( \frac{k}{k_{\textrm{min}}} \right) \right],
\end{eqnarray}
with $\nu$ and $\omega$ denoting
\begin{eqnarray}
\label{eq:sigma}
\nu &=& \underbrace{ \phantom{n} p \phantom{\frac{l}{l_L^5}}}_{(1)} \underbrace{ -\frac{5}{c_L^7} \left( \frac{A}{F} + \frac{B}{F} \right) - \frac{3}{c_L^5} \frac{C}{F} - \frac{2 q}{c_L^5} \widetilde{\mathcal{N}} \frac{D}{F} }_{(2)} \underbrace{ + \frac{ 4 p }{c_L^5} \widehat{\mathcal{N}} \frac{E}{F} }_{(3)}, \\
\label{eq:lambda}
\omega &=& \underbrace{ \frac{2 q}{c_L^5} \frac{D}{F} }_{(2)} \underbrace{ - \frac{4 p}{c_L^5} \frac{E}{F} }_{(3)},
\end{eqnarray}
where the underbraces highlight which terms come from cubic interaction of tensor perturbations with scalar field perturbations '(1)', solid matter perturbations '(2)' and with both of them '(3)', and
\begin{eqnarray}
\label{eq:Ntilde}
\widetilde{\mathcal{N}} & = & N_{\textrm{min}} - \ln \left( 2 c_L \right) + \frac{8}{3} - \gamma_{\textrm{EM}}, \\
\label{eq:Nhat}
\widehat{\mathcal{N}} & = & N_{\textrm{min}} - \ln \left( 1 + c_L \right) + \frac{1}{3} c_L \left( 3 + c_L^2 \right) + \frac{4}{3} - \gamma_{\textrm{EM}}.
\end{eqnarray}
As we can see, both $\widetilde{\mathcal{N}}$ and $\widehat{\mathcal{N}}$ may significantly differ from the minimal number of $e$-folds, but not both of them at the same time. For very small longitudinal sound speed this happens with $\widetilde{\mathcal{N}}$, and for $c_L$ close to one $\widehat{\mathcal{N}}$ no longer approximately equals $N_{\textrm{min}}$.

\vskip 1mm
By analysing orders of magnitude of terms in the tensor-scalar-scalar three-point function (\ref{eq:tss3point}) and the corresponding bispectrum (\ref{eq:tssB}) we can notice differences between three inflationary models in consideration, single-field model, solid inflation, and the combined model. First of all, by reducing the function $F \left( \varphi, X, Y, Z \right)$ to $F \left( \varphi \right)$ we obtain $q = 0$ and our results reduce to results of the single-field inflation. In this case $\nu = p = \epsilon$ and $\omega = 0$, and we find that the tensor-scalar-scalar bispectrum in the squeezed limit and tensor and scalar power spectra satify the consistency relation \cite{maldacena, sreenath} 
\begin{eqnarray}
\label{eq:consistency}
\lim\limits_{K \to 0} {\frac{ \mathcal{B}_{\gamma \zeta \zeta} \left( K, k, k \right) }{ \mathcal{P}_\gamma \left( K \right) \mathcal{P}_\zeta \left( k \right) }} = - \frac{3}{2} + \mathcal{O} \left( \epsilon \right).
\end{eqnarray}
In contrast to this case the solid inflation model, in which $F \left( \varphi, X, Y, Z \right)$ reduces to $F \left( X, Y, Z \right)$ and $p = 0$, breaks the consistency relation \cite{endlich2}. The logarithmic dependence of the bispectrum in solid inflation originates in term with $D/F$ and integral $\mathcal{Y}$ in (\ref{eq:tss3point}) with contribution to parameter $\omega$ of the order $q^2 c_L^{-5}$. This is negligible in comparison to terms in $\nu$ which are of orders $p$, $c_L^{-7} F_Y / F$, $c_L^{-7} \widetilde{F} / F$, $q c_L^{-5}$, and $q^2 c_L^{-5} N_{\textrm{min}}$. In the combined model a new term with $E/F$ and $\mathcal{Y}$ appears in (\ref{eq:tss3point}). Its contribution to $\nu$ is of the order $c_L^{-5} N_{\textrm{min}} \widetilde{F} / F$ and for $\omega$ it is $c_L^{-5} \widetilde{F} / F$. Hence due to interaction between solid matter and scalar field the logarithmic divergence of the tensor-scalar-scalar bispectrum in the limit $\tau_{\textrm{e}} \to 0^{-}$ is considerably enhanced. This means that gravity - solid matter - scalar field interaction is in the squeezed limit much stronger than gravity - solid matter interaction, and the gravity - scalar field interaction is much weaker than both of them.

\subsection{Averaged quadrupole}

We may insert the bispectrum (\ref{eq:tssB}) and power spectra (\ref{eq:powerspectra}) into the formula for the quadrupole (\ref{eq:quadrupole}), but in order to specify it completely we need also the superhorizon tensor perturbation with polarization $\lambda$ and wavevector $\mathbf{K}$, which in principle may be of arbitrary size as long as the perturbation theory is not broken. Of course, the superhorizon tensor perturbation causing the quadrupole asymmetry in the scalar power spectrum is not limited to one specific wavevector and polarization, and the effect must be calculated by summing over them. Moreover, a random character of superhorizon tensor perturbations manifests in randomness of the quadrupole. Physically relevant quantity then can be obtained by averaging over all possible random configurations of tensor perturbations.

\vskip 1mm
Since the inflationary model discussed in this paper predicts nearly Gaussian tensor perturbations, so that $\left< \gamma_{\mathbf{K} i j} \right> \approx 0$, a simple averaging of the quadrupole leads to a negligible quantity which does not contain the most important information about the quadrupole. Therefore, it is conventional to expand the quadrupole part of the scalar bispectrum into the spherical harmonics, $Q^{\lambda}_{ij} \left( K, k \right) k^i k^j / k^2 = \sum_m y^{\lambda}_{2 m} \left( K, k \right) Y_{2 m} \left( \mathbf{k} / k \right)$, and calculate the averaged sum of coefficients of the harmonic expansion squared over coefficients $m$, $\sum_m\left<\left(y^{\lambda}_{2 m}\right)^2\right>$, and then sum it over all polarizations and wavevectors of the superhorizon tensor perturbations. In this way we obtain
\begin{eqnarray}
\label{eq:meanquadrupole0}
\overline{Q^2} \left( k \right) = \frac{16}{15 \pi} \frac{1}{\mathcal{P}_\zeta \left( k \right)^2} \int\limits_{K_{\textrm{min}}}^{k_{\textrm{min}}} K^2 d K \frac{\mathcal{B}_{\gamma\zeta\zeta} \left( K, k, k \right)^2}{\mathcal{P}_\gamma \left( K \right)},
\end{eqnarray}
with integration over superhorizon wavenumbers ranging from wavenumber of the longest mode generated by inflation $K_{\textrm{min}}$ to wavenumber of the longest mode of observational relevance today $k_{\textrm{min}}$, see \cite{dimastrogiovanni}. Note that ratio of two limit wavenumbers is related to the number of $e$-folds during inflation $N$ as $k_{\textrm{min}} / K_{\textrm{min}} = \exp \left( N - N_{\textrm{min}} \right)$.

\vskip 1mm
Finally, using the bispectrum (\ref{eq:tssB}) and power spectra (\ref{eq:powerspectra}) we can evaluate the integral (\ref{eq:meanquadrupole0}) with the result
\begin{eqnarray}
\label{eq:meanquadrupole}
\overline{Q^2}(k)=\frac{12}{5\pi}\frac{H^2}{M_{\textrm{Pl}}^2} \Sigma^2 \left[1+ f \ln\left(\frac{k}{k_{\textrm{min}}}\right)\right]^2\ln\left(\frac{k_{\textrm{min}}}{K_{\textrm{min}}}\right),
\end{eqnarray}
where
\begin{eqnarray}
\Sigma = \frac{c_L^5 \nu}{\epsilon+\left(c_L^5-1\right)p}, \qquad f = \frac{\omega}{\nu}.
\end{eqnarray}
We also have to keep in mind that consistency of the perturbation theory requires smallness of the size of quadrupole. Fortunately, as we can see, it is suppressed by $H / M_{\textrm{Pl}}$. For instance, if the inflation occurs at the scale of grand unified theory, $H / M_{\textrm{Pl}}$ is of the order $10^{-5}$.

\vskip 1mm
As we can see from definitions (\ref{eq:Ntilde}) and (\ref{eq:Nhat}) for generic choice of parameters of the theory both $\widetilde{N}$ and $\widehat{N}$ approximately equal $N_{\textrm{min}}$. By inserting them into (\ref{eq:sigma}) and (\ref{eq:lambda}) we find that $c_L^5 \nu  \approx - \left( 32 / 9 \right) N_{\textrm{min}} \widetilde{F} / F$, and $\omega$ to $\nu$ ratio is approximately $f \approx - N_{\textrm{min}}^{-1}$, which is of the same order as slow-roll parameters. Consequently the averaged quadrupole (\ref{eq:meanquadrupole}) depends on the wavenumber only mildly, $\overline{Q^2} \left( k \right) \approx \overline{Q^2} \left( k_{\textrm{min}} \right) \left(k / k_{\textrm{min}}\right)^{2 f}$, and a similar dependence arises also from next to leading order of slow-roll approximation, which we have omitted. Therefore, for generic choice of parameters of the theory the result (\ref{eq:meanquadrupole}) is correct within the considered approximation only if we set $f$ to zero.

\begin{figure}[!htb]
\sbox0{
\includegraphics[scale=1]{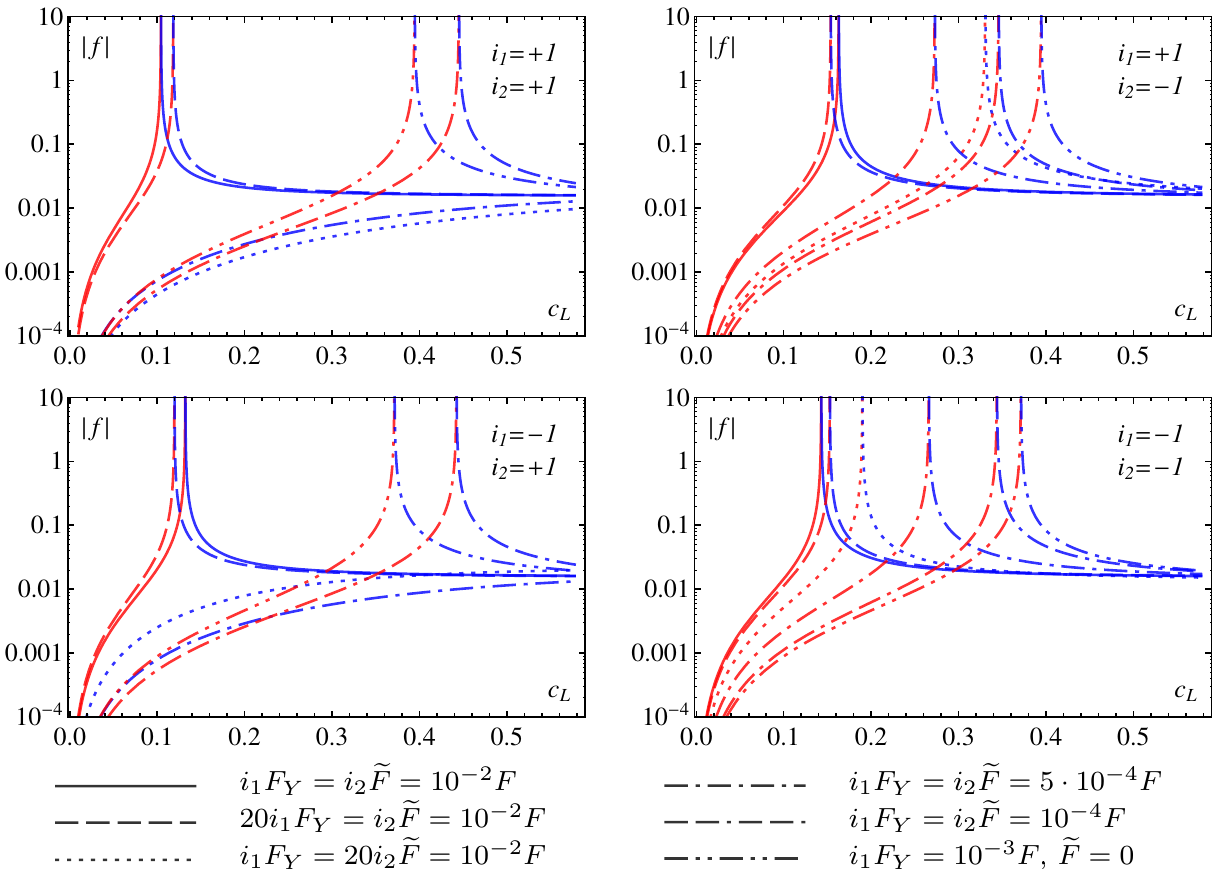}
}
\centering
\begin{minipage}{\wd0}
\usebox0
\linespread{1}
\setlength{\abovecaptionskip}{-8pt plus 0pt minus 0pt}
\caption{{\footnotesize Dependence of absolute value of parameter $f$ on the longitudinal sound speed $c_L$ for $p = q = 10^{-2}$ with other slow-roll parameters set to zero, $N_{\textrm{min}} = 65$, and various values of parameters $F_Y$ and $\widetilde{F}$ with factors $i_1$ and $i_2$ indicating their signs, and four panels corresponding to four possible choices. Parameter $f$ is plotted for $c_L < 1/\sqrt{3}$, red lines are used when $f > 0$, and blue lines correspond to $f < 0$.}}
\label{fig:01}
\end{minipage}
\end{figure}

\vskip 1mm
A more detailed analysis of results reveals that the value of $f$ may be arbitrary. In Fig. \ref{fig:01} we depict its dependence on the longitudinal sound speed $c_L$ with some specific choice of other parameters of the theory. We have set all slow-roll parameters to zero except for $p$ and $q$ set to $10^{-2}$, the transverse sound speed was expressed through (\ref{eq:soundconstraint}) as $c_T^2 = 3 \left( 1 + c_L^2 \right) / 4$ with exact equality, $\widehat{F}$ was replaced by $2 \widetilde{F}$, and for the minimal number of $e$-folds during inflation we have chosen $N_{\textrm{min}} = 65$. We can see that for some choices of $F_Y / F$ and $\widetilde{F} / F$ the $\omega$ to $\nu$ ratio is negative and close to zero while for another choices its absolute value peaks, typically for small values of the sound speed. From definitions of parameters $\nu$ and $\omega$ (\ref{eq:sigma}) and (\ref{eq:lambda}) we can also see that in the leading order of slow-roll approximation $c_L^5 \omega$ is a small constant. Enhancement of parameter $f$ then occurs when $c_L^5 \nu$ approaches zero. This means that enhancement of the averaged quadrupole scale dependence is associated with suppression of its size.

\subsection{Observational consequences}

What qualitatively differentiates our result from results in more standard inflationary models is the logarithmic dependence of the averaged quadrupole on the wavenumber. Another example of an inflationary model with such feature is already mentioned model studied in \cite{rew}. This dependence can be described by relative averaged quadrupole defined by the ratio
\begin{eqnarray}
\label{eq:relativequadrupole}
\frac{\overline{Q^2} \left( k \right)}{ \overline{Q^2} \left( k_{\textrm{min}} \right)} = \left( 1 + f \varkappa \right)^2, \qquad \varkappa = \ln \left( \frac{k}{k_{\textrm{min}}} \right),
\end{eqnarray}
where $\varkappa$ ranges from $\varkappa = 0$ corresponding to $k = k_{\textrm{min}}$ to its maximal value $\varkappa_{\textrm{max}}$ corresponding to wavenumber of the shortest observable perturbations $k_{\textrm{max}}$. For $\varkappa \sim 1$ and smaller the CMB observations are strongly affected by the cosmic variance and dust in our Galaxy and the upper limit on $\varkappa$ given by resolution of the Planck satellite is $\varkappa_{\textrm{max}} \approx 7$.

\vskip 1mm
Concerning the future observations of CMB anisotropies or galaxy surveys with higher resolution, an interesting possibility could be enhancement of the quadrupole asymmetry for very small scale perturbations. This requires $f$ to be of order $\varkappa_{\textrm{max}}^{-1}$ or larger. The dependence of quadrupole on the wavenumber is then in principle detectable by present observations for $f \gtrsim 1/7$. However, currently available analysis of observational data indicate no significant quadrupole asymmetry in scalar power spectrum , and value of the quadrupole $\big(\overline{Q^2}\big)^{1/2}$ for the best fit of Planck data with high $p$-value is of the order $10^{-3}$ \cite{planck18}. So far, any value of parameter $f$ does not contradict the CMB observations.

\vskip 1mm
A more thorough approach to detectability of tensor perturbations with the use of clustering fossils is based on constructing the optimal variance estimator for tensor perturbation amplitude. In \cite{jeong} the variance corresponding to such estimator is defined by sum over wavenumbers and polarizations of tensor perturbations,
\begin{eqnarray}
\label{eq:variance}
\sigma_{\gamma} = \left[ \frac{1}{2} \sum\limits_{\mathbf{K}, \lambda}{ \left( \frac{\mathcal{P}_{\gamma}^{(\textrm{f})} \left( K \right)}{\mathcal{P}_{\lambda}^{(\textrm{n})} \left( K \right)} \right)^2} \right]^{-\frac{1}{2}},
\end{eqnarray}
where $\mathcal{P}_{\gamma}^{(\textrm{f})}$ is the fiducial tensor power spectrum which defines the tensor perturbation amplitude $\mathcal{A}_{\gamma}$ through the tensor power spectrum by relation $\mathcal{P}_{\gamma} \left( K \right) = \mathcal{A}_{\gamma} \mathcal{P}_{\gamma}^{(\textrm{f})} \left( K \right)$. It can be chosen as $\mathcal{P}^{(\textrm{f})} \left( K \right) = K^{n_{\gamma} - 4}$, with tensor spectral index $n_{\gamma}$ close to one for nearly flat spectrum. The noise power spectrum $\mathcal{P}_{\lambda}^{(\textrm{n})} \left( K \right)$ is variance with which a tensor perturbation with wavenumber $K$ and polarization $\lambda$ is measured. For our notations and conventions it can be written as
\begin{eqnarray}
\label{eq:noise}
\mathcal{P}_{\lambda}^{(\textrm{n})} \left( K \right) = 2 V \mathcal{P}_{\gamma} \left( K \right)^2 \left( \sum\limits_{\mathbf{k}}{ \frac{ \mathcal{B}_{\gamma\zeta\zeta} \left( K, k, \left| \mathbf{K} - \mathbf{k} \right| \right)^2 \left| e^{\lambda}_{\mathbf{K} i j} k^i \left( \mathbf{K} - \mathbf{k} \right)^j \right|^2}{k^2 \left( \mathbf{K} - \mathbf{k} \right)^2 \mathcal{P}_{\zeta}^{(\textrm{t})} \left( k \right) \mathcal{P}_{\zeta}^{(\textrm{t})} \left( \left| \mathbf{K} - \mathbf{k} \right| \right)}} \right)^{-1},
\end{eqnarray}
with $V$ denoting the volume of the survey, $V = (2 \pi)^3 / k_{\textrm{min}}^3$, and the total power spectrum $\mathcal{P}^{(\textrm{t})}_{\zeta} \left( k \right)$ containing both signal and noise. Note that these sums can be calculated as integrals, $\sum_{\mathbf{k}} \to V \left( 2 \pi \right)^{-3} \int d^3 k$. Authors of \cite{jeong} applied the variance (\ref{eq:variance}) to the case of single-field model \cite{maldacena}. Under the assumption that the signal-to-noise ratio is $\mathcal{P}_\zeta \left( k \right) / \mathcal{P}^{(\textrm{t})}_\zeta \left( k \right) = 1$ for wavenumbers smaller than the maximal wavenumber of the survey $k_{\textrm{max}}$, and $\mathcal{P}_\zeta \left( k \right) / \mathcal{P}^{(\textrm{t})}_\zeta \left( k \right) = 0$  for $k > k_{\textrm{max}}$, they obtained $\mathcal{P}_{\lambda}^{(\textrm{t})} \left( K \right) = 20 \pi^2 / k_{\textrm{max}}^3$, which for the variance of tensor perturbation aplitude gives $3 \sigma_{\gamma} = 30 \pi \sqrt{3 \pi} \left( k_{\textrm{max}} / k_{\textrm{min}} \right)^{-3}$. Authors studying clustering fossils within solid inflation \cite{dimastrogiovanni, akhshik} obtained possible suppression or enhancement of the right-hand side of this formula depending on choice of parameters of the theory.

\vskip 1mm
For model combining both single-field inflation and solid inflation studied in this paper the generalization of relation for tensor perturbation amplitude variance contains more complicated scale dependence. By inserting power spectra (\ref{eq:powerspectra}) and bispectrum (\ref{eq:tssB}) into (\ref{eq:noise}) and (\ref{eq:variance}) we find $\mathcal{P}_{\lambda}^{(\textrm{t})} \left( K \right) = 20 \pi^2 \mathcal{S} \left( \varkappa_{\textrm{max}} \right) / k_{\textrm{max}}^3$, where
\begin{eqnarray}
\mathcal{S} \left( \varkappa \right) = \frac{1}{\Sigma^2} \left[ 1 + \frac{2}{9} \left( f - 3 \right) f + \left( 2 - \frac{2}{3} f + f \varkappa \right) f \varkappa \right]^{-1},
\end{eqnarray}
and the variance of the tensor perturbation amplitude is
\begin{eqnarray}
\label{eq:variancemodified}
3 \sigma_{\gamma} = 30 \pi \sqrt{3 \pi} \mathcal{S} \left( \varkappa_{\textrm{max}} \right) e^{-3 \varkappa_{\textrm{max}}}.
\end{eqnarray}
These equations were derived using the same approximation for signal-to-noise ratio as in \cite{jeong} and in the limit $k_{\textrm{min}} \ll k_{\textrm{max}}$.

\vskip 1mm
Concerning observations of CMB anisotropies and galaxy surveys other effects must be considered as well. Tensor perturbations affect light propagating from the surface of last scattering or galaxies to the observer, so that their observed positions differs from their real positions \cite{book}. If the tensor-scalar-scalar three-point correlation function obeys consistency relation, this effect cancels the effect of clustering fossils. Therefore, in order to reconcile the theory with observations, parameter $\nu$ defined in (\ref{eq:sigma}) must be replaced by $\nu_{\textrm{obs}}$ in such way that part of the tensor-scalar-scalar bispectrum obeying the consistency relation (\ref{eq:consistency}) is not taken into account,
\begin{eqnarray}
\label{eq:obs}
\nu_{\textrm{obs}} = \nu - \frac{\epsilon + \left( c_L^5 - 1 \right) p}{c_L^5},
\end{eqnarray}
and correspondingly parameters $\Sigma$ and $f$ describing the averaged quadrupole (\ref{eq:meanquadrupole}) must be replaced by $\Sigma_{\textrm{obs}} = c_L^5 \nu_{\textrm{obs}} / \left[ \epsilon + \left( c_L^5 - 1 \right) p \right]$ and $f_{\textrm{obs}} = \omega / \nu_{\textrm{obs}}$. This leads to modification of all results, including the function $\mathcal{S} \left( \varkappa \right)$ appearing in relation for the variance of the tensor perturbation amplitude (\ref{eq:variancemodified}), $\mathcal{S} \left( \varkappa \right) \to \mathcal{S}_{\textrm{obs}} \left( \varkappa \right)$, so that $\sigma_{\gamma}$ must be replaced by $\sigma_{\gamma, \textrm{obs}}$ as well. Galaxy surveys are affected also by growth of structure in later stages of evolution of the universe. This affects both formation of galaxies and propagation of light from them to the observer.

\vskip 1mm
Results of currently available observations must be taken into account as well. In particular, the upper limit on tensor-to-scalar ratio $r$ (\ref{eq:tts}) is $0.064$, and for the absolute value of local nonlinearity parameter $f_{\textrm{NL}}^{(\textrm{local})}$ it is $5.8$. Since the numerical value of the latter experimental restriction was obtained by fitting the Planck data with the use of ordinary local shape of primordial scalar bispectrum, while the scalar bispectrum predicted by both solid inflation and our combined model is more peculiar, the comparison of the nonlinearity parameter (\ref{eq:local}) and the considered observational restriction is only qualitative. Due to the peculiar shape of the bispectrum (\ref{eq:scalarbispectrum}) the parameter $f_{\textrm{NL}}^{(\textrm{local})}$ depends on angle of direction in which the squeezed limit is approached $\theta$. In the context of our work, it is then most reasonable to formulate the observational restriction on the primordial non-Gaussianity in such way, that inequality $\left| f_{\textrm{NL}}^{(\textrm{local})} \right| < 5.8$ is satisfied for every $\theta$ in the interval $0 \leq \theta \leq \pi / 2$.

\begin{figure}[!htb]
\sbox0{
\includegraphics[scale=1]{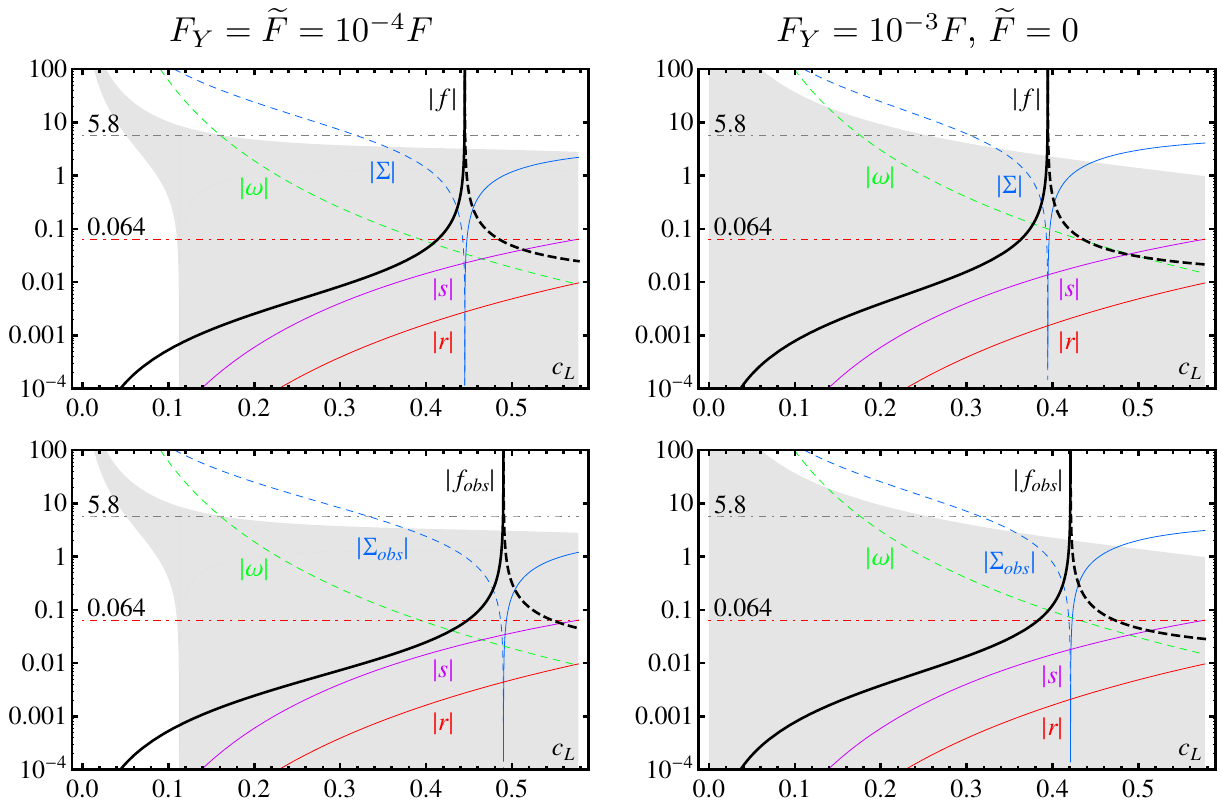}
}
\centering
\begin{minipage}{\wd0}
\usebox0
\linespread{1}
\setlength{\abovecaptionskip}{-8pt plus 0pt minus 0pt}
\caption{{\footnotesize The longitudinal sound speed dependence for absolute values of: parameter $f$ (black line), parameter $\Sigma$ (blue line), parameter $\omega$ (green line), tensor-to-scalar ratio $r$ (red line), nonlinearity parameter $f_{\textrm{NL}}^{(\textrm{local})}$ for every $\theta \in \left[ 0, \pi / 2 \right]$ (gray area) and solid matter perturbations vector-to-scalar ratio $s$ (purple line), in interval $c_L < 1/\sqrt{3}$. All quantities are plotted for $p = q = 10^{-2}$ with other slow-roll parameters set to zero, $N_{\textrm{min}} = 65$, $F_Y = \widetilde{F} = 10^{-4} F$ (upper left panel), $F_Y = 10^{-3} F$ and $\widetilde{F} = 0$ (upper right panel). Dashed lines correspond to negative values. Horizontal lines mark observational restrictions on tensor-to-scalar ratio (red dotted line) and local nonlinearity parameter (gray dotted line). Lower panels copy upper panels with theoretical quantities replaced by observable quantities according to relation (\ref{eq:obs}).}}
\label{fig:02}
\end{minipage}
\end{figure}

\vskip 1mm
With Fig. \ref{fig:01} we have showed that parameter space of the model allows parameter $f$ to be arbitrary. When $F_Y / F, \widetilde{F} / F \sim \epsilon$ its enhancement typically occurs for small longitudinal sound speed, $c_L \sim 10^{-1}$. Unfortunately, parts of the local nonlinearity parameter (\ref{eq:local}) is proportional to $c_L^{-2}$, and in such case it is enhanced beyond compatibility with observations. In order to large values of $\omega$ to $\nu$ ratio to be allowed by observational restrictions, values of parameters $F_Y / F$ and $\widetilde{F} / F$ must be chosen to be much smaller than slow-roll parameters. In Fig. \ref{fig:02} we plot parameters describing the averaged quadrupole (\ref{eq:meanquadrupole}) predicted by the model under consideration as well as parameters on which observational restrictions are imposed as functions of the longitudinal sound speed, for some specific choices of parameters of the theory, in particular, for parameters $F_Y / F$ and $\widetilde{F} / F$ we have set $F_Y / F = \widetilde{F} / F = 10^{-4}$ and $F_Y / F = 10^{-3}$ and $\widetilde{F} / F = 0$. The figure includes also the case when projection effect on light propagating to the observer is taken into account and quantities predicted by the theory must be replaced according to relation (\ref{eq:obs}).

\vskip 1mm
An important observational restriction which we have not considered in this part of the paper is associated with spectral tilt for scalar perturbations. The scalar spectral tilt predicted by the single-field model is $n_{\textrm{s}} - 1 = - 2 \epsilon - \eta$, so that due to the experimental restriction on it two slow-roll parameters $\epsilon$ and $\eta$ are no longer independent. In contrast, the model studied in this paper predicts more complicated scalar spectral tilt (\ref{eq:nsm1}), and even with both slow-roll parameters $p$ and $q$ fixed it can be set to its observed value, as shown in Fig. \ref{fig:00}. The observational restriction on the scalar spectral tilt then does not lead to coupling between slow-roll parameters which we have considered in this section, and for the purpose of interpretation of the obtained results we have been able to choose all parameters appearing in our formulas freely.

\section{Summary}\label{sec:4}

We have calculated quadrupole asymmetry of primordial scalar power spectrum caused by superhorizon tensor perturbations, so called clustering fossils, in model combining both solid inflation and single-field inflation. Object which defines this model is a function of four variables $F \left( \varphi, X, Y, Z \right)$, but freedom in choice of this function is considerably reduced by slow-roll approximation (\ref{eq:Fexpand}), (\ref{eq:tildeF}). Additional restrictions on parameters of the theory are given by observations. In particular, values of tensor-to-scalar ratio $r$ (\ref{eq:tts}) and nonlinearity parameter $f_{\textrm{NL}}^{(\textrm{local})}$ (\ref{eq:local}) cannot be too large \cite{planck18, bicep, planck15}. The scalar spectral tilt imposes another observational restriction, and we have shown that parameter space of the model accommodates regions consistent with it.

\vskip 1mm
In the single-field model consistency relations for correlation functions are satisfied, and quadrupole asymmetry is cancelled out with projection effect of tensor perturbations on light during its propagation to the observer. On the other hand, in the solid inflation model the consistency relations are broken and tensor perturbations leave observable imprint on scalar perturbations. By calculating the tensor-scalar-scalar bispectrum in the squeezed limit within the combined model we revealed that due to interaction between two matter components a logarithmic scale dependence of this bispectrum is not necessarily suppressed by slow-roll parameters. This means that for some choices of parameters of the theory the quadrupole in scalar power spectrum $Q^{(0)}_{ij} k^i k^j / k^2$ with $Q^{(0)}_{ij} \approx$ const. predicted by simpler inflationary models is deformed to $Q_{ij} (k) k^i k^j / k^2$. Therefore, imprint of tensor perturbations on scalar perturbations is scale dependent. More precisely, the scale dependent part of the quadrupole dominates when its scale independent part approaches zero for some choices of free parameters of the model.

\vskip 1mm
In the solid inflation model the logarithmic scale dependence of the quadrupole asymmetry is weak, because it arises only from interaction of the solid with gravity due to nonlinearity of the theory. This weak scale dependence is contaminated by next to leading order of slow-roll approximation, so that it is usually omitted. In the combined model an additional term arises from interaction between scalar field and solid matter with much stronger effect on the asymmetry than solid-gravity interaction. Another inflationary model defined in similar way as the model studied in this paper with non-negligible logarithmic scale dependence of the quadrupole asymmetry is studied by authors of \cite{rew}.

\vskip 1mm
With logarithmically scale dependent quadrupole (\ref{eq:meanquadrupole}) the detectability of primordial tensor perturbations with the use of effect of clustering fossils can be easier with higher angular resolution. This may concern future observations of CMB anisotropies or galaxy surveys. In order to avoid relevance of this effect for angular scales available by current observations the $\omega$ (\ref{eq:lambda}) to $\nu$ (\ref{eq:sigma}) ratio $f$ must be smaller than $1/7$. Of course, the larger its value is the easier the detection of effect of clustering fossils by future observations will be, and since currently available observations do not indicate any significant quadrupole deformation of the primordial scalar power spectrum \cite{planck18}, parameter $f$ is still not restricted from below.

\vskip 1mm
Using the theory from \cite{jeong} we have also calculated the relation between the limit on angular scales available by observations and the variance $\sigma_{\gamma}$ for detection of tensor perturbations with the use of the effect of clustering fossils (\ref{eq:variancemodified}). This relation is in the combined model more complicated than in both single-field model and solid inflation.

\section*{Acknowledgement}
The work was supported by the grant VEGA 1/0985/16.

\appendix
\renewcommand{\thesection}{\Alph{section}}

\section{Slow-roll restrictions on matter Lagrangian}\label{app:a}

Restrictions on the function $F \left( \varphi, X, Y, Z \right)$ determining the matter Lagrangian of our model given by relations (\ref{eq:Fexpand}) and (\ref{eq:tildeF}), which we have used throughout the text, follow from slow-roll approximation. Relation (\ref{eq:Fexpand}) is valid also in the case with arbitrary size of slow-roll parameters appearing in it, so that its applicability is not reduced to nearly de Sitter universe. This is true also for all relations appearing in this appendix where we derive coefficients of perturbative expansion of the function $F$.

\paragraph{(Zeroth order)}

Starting expansion of the function $F \left( \varphi, X, Y, Z \right)$ with the unperturbed case, we simply rewrite the Friedmann equation (\ref{eq:friedmann1}) with the use of definition of parameter $p$ (\ref{eq:epsilon}) as
\begin{eqnarray}
F = -3 M_{\textrm{Pl}}^2 H^2 + \frac{1}{2} \dot{\varphi}^2 = M_{\textrm{Pl}}^2 H^2 \left( - 3 + p \right).
\end{eqnarray}

\paragraph{(First order)}

First partial derivatives of the function $F$ explicitly appear in equations (\ref{eq:friedmann2}) and (\ref{eq:kleingordon}). The latter one contains second time derivative of the scalar field, which can be rewritten with the use of definitions (\ref{eq:slowroll}) and (\ref{eq:epsilon}) as
\begin{eqnarray}
\label{eq:F0}
\ddot{\varphi} = \frac{d}{dt} \dot{\varphi} = \pm \sqrt{2} M_{\textrm{Pl}} \frac{d}{dt} H \sqrt{p} = \pm \sqrt{2} M_{\textrm{Pl}} H^2 \sqrt{p} \left( \frac{1}{2} \eta_p - \epsilon \right).
\end{eqnarray}
By inserting this expression into (\ref{eq:kleingordon}) we find
\begin{eqnarray}
\label{eq:Fvarphi}
F_{\varphi} = \pm \sqrt{2} M_{\textrm{Pl}} H^2 \sqrt{p} \left( 3 - \epsilon + \frac{1}{2} \eta_p \right).
\end{eqnarray}
Finally, using (\ref{eq:slowroll}) and (\ref{eq:epsilon}) we can simply rewrite (\ref{eq:friedmann2}) as
\begin{eqnarray}
\label{eq:FX}
X F_X = 3 M_{\textrm{Pl}}^2 H^2 \left( p - \epsilon \right) = - 3 M_{\textrm{Pl}}^2 H^2 Q,
\end{eqnarray}
where $X$ on the left-hand side corresponds to the unperturbed case. So far there are no restrictions on $F_Y$ and $F_Z$, because $\delta Y$ and $\delta Z$ are of the second order

\paragraph{(Second order)}

By differentiating first partial derivatives of the function $F$ we obtain
\begin{eqnarray}
\dot{F}_{\varphi} = F_{\varphi \varphi} \dot{\varphi} + F_{\varphi X} \dot{X},
\qquad
\dot{F}_X = F_{\varphi X} \dot{\varphi} + F_{XX} \dot{X}.
\end{eqnarray}
The left-hand sides of these equations can be calculated by differentiating (\ref{eq:Fvarphi}) and (\ref{eq:FX}) and using definitions of slow-roll parameters. On the right-hand sides, $\dot{\varphi}$ can be rewritten in terms of $p$ as in (\ref{eq:F0}), and $\dot{X} = - 2 H X$. In this way we obtain
\begin{eqnarray}
\label{eq:pom1}
& & \pm \sqrt{2} M_{\textrm{Pl}} \sqrt{p} F_{\varphi \varphi} - 2 X F_{\varphi X} = \\
& & = \pm \sqrt{2} M_{\textrm{Pl}} H^2 \sqrt{p} \left[ \left( - 2 \epsilon + \frac{1}{2} \eta_p \right) \left( 3 - \epsilon + \frac{1}{2} \eta_p \right) + \frac{1}{2} \eta_p \eta^{(2)}_p - \epsilon \eta \right], \nonumber\\
\label{eq:pom2}
& & \pm \sqrt{2} M_{\textrm{Pl}} \sqrt{p} X F_{\varphi X} - 2 X^2 F_{XX} = - 6 M_{\textrm{Pl}}^2 H^2 Q \left( 1 - \epsilon + \frac{1}{2} \eta_Q \right).
\end{eqnarray}
Unfortunately, here we have only two equations for three variables, $F_{\varphi \varphi}$, $F_{\varphi X}$ and $F_{XX}$. Therefore, we have to use also expressions for sound speeds (\ref{eq:soundspeeds}), so that we increase number of equations by two with only one additional variable $F_Y + F_Z$. In order to avoid superluminal or imaginary sound speeds, partial derivatives of the function $F$ must be restricted,
\begin{eqnarray}
\left| X^2 F_{XX} \right| < 2 \left| X F_X \right|, \qquad \left| F_Y + F_Z \right| < \frac{3}{2} \left| X F_X \right|.
\end{eqnarray}
Specifically, parameters $F_{XX}$ and $F_Y + F_Z$ can be expressed in terms of sound speeds as
\begin{eqnarray}
\label{eq:ddF1}
\label{eq:0001}
X^2 F_{XX} = 6 M_{\textrm{Pl}}^2 \mathcal{C}_1 H^2 Q, \qquad F_Y + F_Z = \frac{9}{2} M_{\textrm{Pl}}^2 \mathcal{C}_2 H^2 Q,
\end{eqnarray}
where $\mathcal{C}_1$ and $\mathcal{C}_2$ are defined in (\ref{eq:cecka}). Finally, by inserting $F_{XX}$ into (\ref{eq:pom1}) and (\ref{eq:pom2}) we find
\begin{eqnarray}
\label{eq:ddF2}
X F_{\varphi X} = \pm 9 \sqrt{2} M_{\textrm{Pl}} \mathcal{C}_3 H^2 \frac{Q}{\sqrt{p}}, \qquad F_{\varphi \varphi} = 18 \mathcal{C}_4 H^2 \frac{Q}{p},
\end{eqnarray}
with $\mathcal{C}_3$ and $\mathcal{C}_4$ defined in (\ref{eq:cecka}) as well. Of course, one of parameters $F_Y$ and $F_Z$ may be arbitrary, as long as their sum is given by the second relation in (\ref{eq:0001}).

\paragraph{(Third order)}

Coefficients of perturbative expansion of function $F$ beyond second order cannot be fully specified in terms of slow-roll parameters and sound speeds. However, by differentiating (\ref{eq:ddF1}) and (\ref{eq:ddF2}) and using the same procedure as above we find that these coefficients are restricted by conditions
\begin{eqnarray}
\pm \sqrt{2} M_{\textrm{Pl}} \sqrt{p} X^2 F_{\varphi X X} - 2 X^3 F_{X X X} & = & 6 M_{\textrm{Pl}}^2 \mathcal{C}_1 H^2 Q \left( 4 + d_1 \right), \\
\label{eq:thesecondrelation}
\pm \sqrt{2} M_{\textrm{Pl}} \sqrt{p} \left( F_{\varphi Y} + F_{\varphi Z} \right) - 2 X \left( F_{X Y} + F_{X Z} \right) & = & \frac{9}{2} M_{\textrm{Pl}}^2 \mathcal{C}_2 H^2 Q d_2, \\
M_{\textrm{Pl}} p X F_{\varphi \varphi X} \mp \sqrt{2 p} X^2 F_{\varphi X X} & = & 9 M_{\textrm{Pl}} \mathcal{C}_3 H^2 Q \left( 2 + d_3 \right), \\
\pm \sqrt{2} M_{\textrm{Pl}} p^{\frac{3}{2}} F_{\varphi \varphi \varphi} - 2 p X F_{\varphi \varphi X} & = & 18 \mathcal{C}_4 H^2 Q d_4,
\end{eqnarray}
where $ d_1 = - 2 \epsilon + \eta_Q + \dot{\mathcal{C}}_1 / ( H \mathcal{C}_1 ) $, $ d_2 = - 2 \epsilon + \eta_Q + \dot{\mathcal{C}}_2 / ( H \mathcal{C}_2 ) $, $ d_3 = - 2 \epsilon - \eta_p/2 + \eta_Q + \dot{\mathcal{C}}_3 / ( H \mathcal{C}_3 ) $, $ d_4 = - 2 \epsilon - \eta_p + \eta_Q + \dot{\mathcal{C}}_4 / ( H \mathcal{C}_4 ) $, and $\dot{\mathcal{C}_1}$, ... $\dot{\mathcal{C}_4}$ can be expressed in terms of other slow-roll parameters including $\eta_L$ and $\eta_T$ measuring rate of change of the sound speeds.
Equation (\ref{eq:thesecondrelation}) was considered when the parameter $\widetilde{F}$ and its relation with $\widehat{F}$ was defined by (\ref{eq:tildeF}).

\section{Lam\'e parameters}\label{app:b}

In Newtonian mechanics of continuum the equation of state determining dependence of energy density of a homogeneous and isotropic solid on its deformations is parametrised by two Lam\'e parameters $\lambda$ and $\mu$ as \cite{ll}
\begin{eqnarray}
\label{eq:equationofstate}
\rho_{\textrm{sol}} = \frac{1}{2} \lambda \left( \textrm{Tr} \phantom{|} u \right)^2 + \mu \textrm{Tr} \phantom{|} u^2,
\end{eqnarray}
where a small deformation of solid is encoded by the strain tensor $u_{ij} = \left( \pi_{i,j} + \pi_{j,i} \right) / 2$, with $\pi_i$ denoting components of the displacement vector defined in the same way as a sum $\delta^{Ii} \rho_{,i} + \pi^I_{\perp}$ appearing in (\ref{eq:displacement}).

\vskip 1mm
For deformations of arbitrary size the strain tensor is replaced by the body metric defined as push-forward of the Euclidean metric with respect to mapping between space coordinates $x^i$ and body coordinates $\phi^i$ multiplied by factor $1/2$,
\begin{eqnarray}
u_{ij} = \frac{1}{2} \left( \pi_{i,j} + \pi_{j,i} + \pi_{i,k} \pi_{j,k} \right),
\end{eqnarray}
and a relativistic generalization of elasticity is built by defining the body metric $B$ as a push-forward of the space-time metric (\ref{eq:bodymetric}). The relativistic equation of state leading to (\ref{eq:equationofstate}) in the Newtonian limit can be chosen as
\begin{eqnarray}
\label{eq:relequationofstate}
\rho_{\textrm{sol}} = \frac{n}{\overline{n}} \rho_{\overline{n}}, \qquad \rho_{\overline{n}} = \overline{\rho} - P \frac{\delta V} {\overline{V}} + \frac{1}{8} \lambda \left[ \delta B \right]^2 + \frac{1}{4} \mu \left[ \delta B^2 \right],
\end{eqnarray}
where $\rho_{\overline{n}}$ is energy density associated with concentration of solid matter elements $\overline{n}$, the pressure $P$ is defined by relative change in volume $\delta V / \overline{V}$, and square brackets denote traces of the body metric perturbation $\delta B_{IJ} = B_{IJ} - B^{(0)}_{IJ}$ defined by raising and lowering the indices with respect to the background body metric, $\left[ \delta B \right] = B^{(0) IJ} \delta B_{IJ}$, $\left[ \delta B^2 \right] = B^{(0) IK} B^{(0) JL} \delta B_{IJ}$  $\delta B_{KL}$. The pressure and Lam\'e parameters can be considered as coefficients of expansion of the equation of state up to the second order in body metric perturbation, and some of works with this approach to relativistic elasticity, see \cite{bucher, karlovini}, use different definitions for them. Here we follow conventions used in \cite{skovran}. By expanding the relative concentration and volume up to the second order,
\begin{eqnarray}
\frac{n}{\overline{n}} = 1 - \frac{1}{2} \left[ \delta B \right] + \frac{1}{8} \left[ \delta B \right]^2 + \frac{1}{4} \left[ \delta B^2 \right], \qquad \frac{V}{\overline{V}} = 1 + \frac{1}{2} \left[ \delta B \right] + \frac{1}{8} \left[ \delta B \right]^2 - \frac{1}{4} \left[ \delta B^2 \right],
\end{eqnarray}
we can rewrite the equation of state (\ref{eq:relequationofstate}) as
\begin{eqnarray}
\label{eq:relequationofstate2}
\rho_{\textrm{sol}} = \overline{\rho} \left( 1 - \frac{1}{2} \left( w + 1 \right) \left[ \delta B \right] + \frac{1}{8} \left( \lambda_0 + w + 1 \right) \left[ \delta B \right]^2 + \frac{1}{4} \left( \mu_0 + w + 1 \right) \left[ \delta B^2 \right] \right),
\end{eqnarray}
where $w = P / \overline{\rho}$ is the pressure to energy density ratio, and $\lambda_0 = \lambda / \overline{\rho}$ and $\mu_0 = \mu / \overline{\rho}$ are dimensionless Lam\'e parameters.

\vskip 1mm
In this paper we worked with a more general equation of state,
\begin{eqnarray}
\rho_{\textrm{sol}} = \mathcal{E} \left( X, Y, Z \right),
\end{eqnarray}
with quantities $X, Y, Z$ defined by traces of the body metric (\ref{eq:invariants}). Their perturbations can be expanded in terms of the body metric perturbation up to the second order as
\begin{eqnarray}
\delta X = \frac{1}{a^2} \left( - \left[ \delta B \right] + \left[ \delta B^2 \right] \right), \qquad \delta Y = \delta Z = -\frac{1}{27} \left[ \delta B \right]^2 + \frac{1}{9} \left[ \delta B^2 \right],
\end{eqnarray}
and the perturbative expansion of the general equation of state is
\begin{eqnarray}
\mathcal{E} = \overline{\mathcal{E}} - \frac{\mathcal{E}_X}{a^2} \left[ \delta B \right] + \left( \frac{\mathcal{E}_{XX}}{2a^4} - \frac{\mathcal{E}_Y + \mathcal{E}_Z}{27} \right) \left[ \delta B \right]^2 + \left( \frac{\mathcal{E}_X}{a^2} + \frac{\mathcal{E}_Y + \mathcal{E}_Z}{9} \right) \left[ \delta B^2 \right].
\end{eqnarray}
By comparing this relation with (\ref{eq:relequationofstate2}) we find the dimensionless Lam\'e parameters of the solid as well as its pressure to energy ratio,
\begin{eqnarray}
& & \lambda_0 = - \frac{2}{a^2} \frac{\mathcal{E}_X}{\overline{\mathcal{E}}} + \frac{4}{a^4} \frac{\mathcal{E}_{XX}}{\overline{\mathcal{E}}} - \frac{8}{27} \frac{\mathcal{E}_Y + \mathcal{E}_Z}{\overline{\mathcal{E}}}, \\
& & \mu_0 = \frac{2}{a^2} \frac{\mathcal{E}_X}{\overline{\mathcal{E}}} + \frac{4}{9} \frac{\mathcal{E}_Y + \mathcal{E}_Z}{\overline{\mathcal{E}}}, \qquad w = -1 + \frac{2}{a^2} \frac{\mathcal{E}_X}{\overline{\mathcal{E}}}. \nonumber
\end{eqnarray}
The sound speeds (\ref{eq:soundspeeds}) with function $F \left( \varphi, X, Y, Z \right)$ replaced by function $\mathcal{E} \left( X, Y, Z \right)$ can be rewritten as
\begin{eqnarray}
c_T^2 = \frac{\mu_0}{w + 1}, \qquad c_L^2 = \frac{\lambda_0 + 2 \mu_0}{w + 1},
\end{eqnarray}
so that they depend on Lam\'e parameters in the same way as sound speeds in Newtonian elasticity. Note also that the dimensionless compressional modulus,
\begin{eqnarray}
K_0 = \lambda_0 + \frac{2}{3} \mu_0 = - \frac{2}{3 a^2} \frac{\mathcal{E}_X}{\overline{\mathcal{E}}} + \frac{4}{a^4} \frac{\mathcal{E}_{XX}}{\overline{\mathcal{E}}},
\end{eqnarray}
depends only on $X$ and derivatives of function $\mathcal{E}$ with respect to it.

\section{Quadrupole invariance of angular power spectrum}\label{app:x}

The observed anisotropies in CMB temperature are given by evolution of primordial perturbations generated during inflation and line-of-sight integral describing effects influencing photons during their propagation to the observer. The line-of-sight integral is considerably simplified if we take into account only the linearised perturbation theory and only scalar perturbations within it, assume that photons were in local equilibrium during recombination, and that they scattered for the last time at any moment $\tau_{\textrm{L}}$ with probability distribution given by their mean free time \cite{weinberg}. Perturbation in the CMB temperature $\delta T$ as a function of direction $\mathbf{n}$ from which the photons are arriving to the observer at time $\tau_0$ then can be estimated as
\begin{eqnarray}
\label{eq:fluctuation}
\delta T \left( \mathbf{n} \right) = \int\limits_0^{\tau_0} d \tau_{\textrm{L}} \int\limits_{\tau_{\textrm{L}}}^{\tau_0} d \tau \int \frac{d^3 k}{\left( 2 \pi \right)^3} \widehat{\mathcal{T}} \left( \tau_{\textrm{L}}, \tau, k \right) \Phi_{\mathbf{k}} e^{i \mathbf{k} \cdot \mathbf{n} \left( \tau - \tau_0 \right)},
\end{eqnarray}
where $\Phi_{\mathbf{k}}$ is Fourier mode of the primordial Newtonian potential, and
\begin{eqnarray}
\widehat{\mathcal{T}} \left( \tau_{\textrm{L}}, \tau, k \right) = \left[ \left( \mathcal{T}_{\textrm{SW}} \left( k \right) + \mathcal{T}_{\textrm{D}} \left( k \right) \frac{\partial}{\partial \tau_0} \right) \delta \left( \tau - \tau_{\textrm{L}} \right) + \mathcal{T}_{\textrm{ISW}} \left( \tau, k \right) \right] P_{\textrm{V}} \left( \tau_{\textrm{L}} \right),
\end{eqnarray}
with $\mathcal{T}_{\textrm{SW}}$ denoting part of the transfer-function describing Sachs--Wolfe effect, $\mathcal{T}_{\textrm{D}}$ is for Doppler effect, $\mathcal{T}_{\textrm{ISW}}$ describes integrated Sachs--Wolfe effect, and $P_{\textrm{V}}$ is visibility function \cite{cmbslow}. Note that the transfer-function $\widehat{\mathcal{T}} \left( \tau_{\textrm{L}}, \tau, k \right)$ depends only on size of the wavenumber $k$, because equations governing evolution of perturbations within the standard $\Lambda$CDM model are isotropic.

\vskip 1mm
The CMB temperature fluctuations (\ref{eq:fluctuation}) can be expanded into spherical harmonics with coefficients of the expansion
\begin{eqnarray}
\label{eq:alm}
a_{l m} = \int d^2 n \delta T \left( \mathbf{n} \right) Y^{*}_{l m} \left( \mathbf{n} \right) = 4 \pi \left( - i \right)^l \int d^2 \tau \frac{d^3 k}{\left( 2 \pi \right)^3} j_{l} \left( k \Delta \right) \widehat{\mathcal{T}} \left( \tau_{\textrm{L}}, \tau, k \right) \Phi_{\mathbf{k}} Y^{*}_{lm} \left( \mathbf{q} \right),
\end{eqnarray}
where $\mathbf{q} = \mathbf{k} / k$, $\Delta = \tau_0 - \tau$, and $\int d^2 \tau$ is a shorthand for $\int_0^{\tau_0} d \tau_{\textrm{L}} \int_{\tau_{\textrm{L}}}^{\tau_0} d \tau$. We have expanded plane waves into spherical harmonics with the use of spherical Bessel functions of the first kind $j_l$, and then we have made use of the fact that spherical harmonics are orthonormal. The angular power spectrum coefficients predicted by the theory are defined by averaging of absolute values of coefficients (\ref{eq:alm}) squared, $C_l = \sum_m \left< \left| a_{l m} \right|^2 \right> / \left( 2 l + 1 \right)$, and the direct calculation leads to
\begin{eqnarray}
\label{eq:cl1}
C_l & = & \frac{\left( 4 \pi \right)^2 \left( - 1 \right)^l}{ 2 l + 1 } \int d^2 \tau \frac{d^3 k}{\left( 2 \pi \right)^3} d^2 \tau^{\prime} \frac{d^3 k^{\prime}}{\left( 2 \pi \right)^3} j_l \left( k \Delta \right) j_l \left( k^{\prime} \Delta^{\prime} \right) \\
& & \widehat{\mathcal{T}} \left( \tau_{\textrm{L}}, \tau, k \right) \widehat{\mathcal{T}} \left( \tau_{\textrm{L}}^{\prime}, \tau^{\prime}, k^{\prime} \right) \left< \Phi_{\mathbf{k}} \Phi_{\mathbf{k}^{\prime}} \right> \sum\limits_m Y^{*}_{lm} \left( \mathbf{q} \right) Y_{lm} \left( \mathbf{q}^{\prime} \right). \nonumber
\end{eqnarray}
Now we assume the two-point correlation function of the primordial Newtonian potential to be of the same form as (\ref{eq:toquad}),
\begin{eqnarray}
\left< \Phi_{\mathbf{k}} \Phi_{\mathbf{k}^{\prime}} \right> = \left( 2 \pi \right)^3 \delta^{(3)} \left( \mathbf{k} + \mathbf{k}^{\prime} \right) \mathcal{P}^{(0)} \left( k \right) \left( 1 + Q_{ij} \left( k \right) q^i q^j \right).
\end{eqnarray}
By inserting it into (\ref{eq:cl1}), integrating over $\mathbf{k}^{\prime}$-space, and using the Uns\"old's theorem, $4 \pi $ $\sum_m$ $Y^{*}_{lm} \left( \mathbf{q} \right)$ $Y_{lm} \left( \mathbf{q} \right) =$ $2 l + 1$, we find
\begin{eqnarray}
C_l & = & \frac{2}{\pi} \int d^2 \tau d^2 \tau^{\prime} k^2 dk j_l \left( k \Delta \right) j_l \left( k \Delta^{\prime} \right) \\
& & \widehat{\mathcal{T}} \left( \tau_{\textrm{L}}, \tau, k \right) \widehat{\mathcal{T}} \left( \tau_{\textrm{L}}^{\prime}, \tau^{\prime}, k \right) \mathcal{P}^{(0)} \left( k \right) \left( 1 + Q_{ij} \left( k \right) \int \frac{d^2 q}{4 \pi} q^i q^j \right). \nonumber
\end{eqnarray}
The integral in brackets is
\begin{eqnarray}
Q_{ij} \left( k \right) \int \frac{d^2 q}{4 \pi} q^i q^j = \frac{1}{3} \textrm{Tr} \phantom{|} Q \left( k \right) = 0,
\end{eqnarray}
so that the angular power spectrum coefficients do not depend on the quadrupole.

{\setstretch{1.0}

}

\end{document}